\documentclass[sigconf,nonacm]{acmart}
\usepackage{dcolumn}
\AtBeginDocument{%
  \providecommand\BibTeX{{%
    \normalfont B\kern-0.5em{\scshape i\kern-0.25em b}\kern-0.8em\TeX}}}

\setcopyright{acmcopyright}
\copyrightyear{2022}
\acmYear{2022}
\acmDOI{XXXXXXX.XXXXXXX}

\begin{document}

\title{Birdwatch: Crowd Wisdom and Bridging Algorithms can Inform Understanding and Reduce the Spread of Misinformation}

\author{Stefan Wojcik}
\affiliation{%
 \institution{Twitter Cortex}}

\author{Sophie Hilgard}
\affiliation{%
 \institution{Twitter Cortex}}

\author{Nick Judd}
\affiliation{%
 \institution{Twitter Research}}

\author{Delia Mocanu}
\affiliation{%
 \institution{Twitter Cortex}}

\author{Stephen Ragain}
\affiliation{%
 \institution{Twitter Cortex}}

\author{M.B. Fallin Hunzaker}
\affiliation{%
 \institution{Twitter Research}}

\author{Keith Coleman}
\affiliation{%
 \institution{Twitter Product}}

\author{Jay Baxter}
\authornote{Corresponding Author: birdwatch.jay@gmail.com}

\affiliation{%
  \institution{Twitter Cortex}}

\renewcommand{\shortauthors}{Wojcik et al.}

\begin{abstract}
  We present an approach for selecting objectively informative and subjectively helpful annotations to social media posts. We draw on data from on an online environment where contributors annotate misinformation and simultaneously rate the contributions of others. Our algorithm uses a matrix-factorization (MF) based approach to identify annotations that appeal broadly across heterogeneous user groups --- sometimes referred to as ``bridging-based ranking.'' We pair these data with a survey experiment in which individuals are randomly assigned to see annotations to posts.
  We find that annotations selected by the algorithm improve key indicators compared with overall average and crowd-generated baselines. Further, when deployed on Twitter, people who saw annotations selected through this bridging-based approach were significantly less likely to reshare social media posts than those who did not see the annotations.
\end{abstract}
\maketitle



\section{Introduction}
Social media remains a critical part of how Americans consume news \cite{walker2021news}. Social media companies seek to meet this need by surfacing credible news content from diverse voices \cite{coleman2021introducing, clegg_2022, geary_2021}. However, misinformation presents a lingering challenge. Especially for polarizing topics like news or politics, surfacing content found credible by a broad audience remains a challenge \cite{kozyreva2022free}.

Twitter's Birdwatch feature \cite{coleman2021introducing} is a community-driven approach to tackle misinformation on the platform. Birdwatch allows users to collaboratively add ``notes,'' or contextual ``annotations,''  to Tweets. Birdwatch contributors may seek to call out errors or omissions in a Tweet they identify as misleading, or simply to add additional context that might be helpful for understanding the issues raised in the Tweet. Other contributors rate notes as ``Helpful,'' ``Not helpful,'' or ``Somewhat helpful,'' based on whether the note provides helpful context or clarification about the Tweet or the issues it raises. 

The Birdwatch project takes on two key goals. The first is to surface high-quality, user-generated notes that are objectively informative. In this paper, ``objectively informative''summary notes are those that causally improve readers' understanding of core claims in potentially misleading Tweets. The second goal is to surface notes that are perceived as helpful by a broad cross-section of people on Twitter.

 Identifying notes that satisfy both aims is a challenge. For instance, some well-sourced notes may be seen as unhelpful because they are poorly written, or because they use language that may be perceived as biased or argumentative. For instance, someone might feel a note is combative — or simply hard to read — and take that as a cue to disregard the information it contains, rather than consider the note’s salient, if ineptly presented, information.
 
 Similarly, notes with weak sourcing, or without a strong factual basis, may appeal to people by invoking taken-for-granted ideas or assumptions. For example, large groups of people on Twitter might agree with, and rate as helpful, notes that are misleading or non-informative but also consistent with their prior beliefs \cite{taber2006motivated}. A core challenge for Birdwatch, then, is to identify notes which not only contain accurate, high quality information, but are also written in a way that is likely to resonate with broad audiences, not just those who are already inclined to agree.

We present an algorithm to identify which notes are informative and helpful based on the user-generated notes themselves and the history of user-generated ratings for each note. Using these inputs, we seek to overcome two obstacles to our objectives. First, ratings are themselves a function of not only a note's latent properties (e.g. quality, tone, bias), but also of how raters react to the note, given each rater's prior beliefs. Second, we have no prior information about each rater's prior beliefs, each note's latent properties, or how these attributes interact in the process that generates individual ratings. We develop a matrix factorization (MF) method from the rater-note matrix in Birdwatch that captures the baseline propensity of raters to rate notes as ``helpful'' and a rater's ``viewpoint'' based on their past ratings of similar notes.

We show that notes selected by this algorithm are found subjectively helpful by individuals from diverse political viewpoints, and present causal evidence that notes also inform readers' understanding of social media posts. We base our results on data from a limited rollout of the Birdwatch feature that was made available to a subset of users within the US in 2021–2022 and multiple waves of a survey experiment conducted with representative samples of Twitter users. We measure “helpfulness” by showing survey respondents a randomly selected note and asking their opinion of the note. We measure informativeness by estimating the effect of being exposed to a note on the probability of agreeing with different statements summarizing the claims of a potentially misleading Tweet. We find effects on informativeness that are comparable to those uncovered in recent large-scale evaluations of journalistic fact-checks. The implication, which we will discuss, is that it is possible to identify a set of crowd-sourced social media annotations that are, on average, strongly informative.

The measurement strategy we use to assess informativeness is in line with recent related literature on journalistic fact-checks, but it is an advance relative to a common measurement strategy in studies of social media (i.e., measuring perceptions of content on true-false accuracy scales). We show that simply measuring perceptions of social media post accuracy fails to capture substantial effects on the informativeness of additional context. This helps resolve a core confusion in the literature around existing measures of annotation quality, which sometimes assumes high correspondence between the concepts of objective informativeness of misinformation annotations and perceived accuracy of social media posts on true/false scales.

Finally, we show that annotations selected by this algorithm inform people’s social media sharing behaviors. People choose to share or “Like” posts significantly less when they see an annotation, relative to those who do not see annotations.

The development of our matrix factorization algorithm was the result of focusing on the following three research questions motivated by project goals and the literature:

\textbf{RQ1:} Can we select a set of Birdwatch notes that both inform understanding (decrease propensity to agree with a potentially misleading claim) and are seen as helpful by a diverse population of users (in particular, users with diverse self-reported political affiliations)? Does algorithmic selection achieve these better than a supermajority voting baseline? 

\textbf{RQ2:} How closely related are appraisals of the accuracy of a Tweet and agreement with a Tweet's claims? Can appraisals of the accuracy of a Tweet reliably be used as proxies for agreement with a Tweet's claims? 

\textbf{RQ3:} Can crowd-generated Tweet annotations reduce the spread of potentially misleading information on Twitter? Does exposure to Tweet annotations reduce retweeting and "Liking" behavior compared to users who do not see annotations?  

\section{Related Work}

Two closely related streams of misinformation research inform the present work. The first concerns whether  journalistic ``fact-checks'' can correct misunderstandings caused by misleading claims, including or especially when those claims align with the reader's prior beliefs \cite{nyhan2020facts, nyhan2021backfire}. The second concerns annotations on misleading claims in the specific context of crowds on social media. In the first, 
the central preoccupations are (1) any causal effect of corrective information on misperceptions and (2) the processes and social structures that explain observed heterogeneity in such effects. In the second,
common questions include (1) whether annotations change social media users' perceived accuracy of misleading posts and (2) whether social media-based crowds might be an effective means to scale up the production of such annotations.

\subsection{Fact-checks and User Perceptions}

Fact-checks produced by fact-checking organizations reduce the extent to which people agree with misleading claims, an effect that has recently been shown to generalize across national contexts and to be durable over time \cite{porter2022political}. 

However, even with additional context or corrective statements in hand, social media companies must decide whether and how to apply context-focused content into potentially misleading posts. Existing interventions take many forms, and many studies have demonstrated that informational cues such as warning labels/flags \cite{clayton2020real}, prebunking \cite{lewandowsky2021countering}, or providing access to fact-checks \cite{nyhan2020taking} all impact how people on social media interpret what they see online. 

Labels in particular may be politically polarizing to users, and sometimes only certain subsets of users trust them. In survey research in the U.S., studies have found a reduced level of trust, self-reported efficacy, and overall support of misinformation warning labels among self-identified Republicans relative to Democrats and independents. In one study, 15\% fewer Republicans than Democrats reported that warning labels made them less likely to trust the contents of a post \cite{straub2022americans}. Similarly, \citet{walker2019republicans} show using survey data that about 70\% of Republicans think fact-checkers will favor a particular side. \citet{mitchell2021more} analyze data from a  nationally-representative survey of US adults, and find a widening gap between Democrats and Republicans regarding whether tech companies should take steps to restrict false information on their platforms.

\subsection{Crowd-based Annotations}

Because a large number of misleading claims may be introduced at the same time, and because each claim may be repeated many times as they quickly spread through dense networks \cite{vosoughi2018spread}, it is challenging for fact-checks created by a limited set of experts to keep pace with misinformation in real-world applications \cite{rodrigo2020critics}. Further, research suggests that public trust in companies and government to moderate content is lacking \cite{stocking_etal_role}. To combat misinformation at scale, new approaches seek community-based solutions to the spread of misinformation \cite{silverman2019helping, allen2021scaling}. However, the participation of misinformed and malicious actors can undermine this process, providing an outlet for misinformation to spread  \cite{lazer2018science}. 

Many previous works have shown that layperson crowds can identify and objectively label misinformation or low-quality sources \cite{pennycook2019fighting, golub2010naive, woolley2010evidence, resnick2021informed, allen2021scaling}. Several such studies focus on appraisals of the accuracy of a headline or a brief block of text. For instance, one recent study found that small, politically balanced groups of crowdworkers produce accuracy ratings that predict whether a majority of fact-checkers would agree whether any given headline was true or false \cite{allen2021scaling}.

A similarly rich body of work seeks to understand what drives crowdworkers toward (or away from) accurate assessments of news. Some studies identify politically conservative \cite{roitero2020can, bhuiyan2020investigating, barbera2020crowdsourcing} or highly-polarized \cite{roitero2020covid} crowdworkers as generally less accurate, while others give reason to believe that it may be possible to achieve broad consensus even among strongly polarized crowdworkers. For example, crowdworkers on both sides of the political spectrum agree when evaluating news \emph{source quality} \cite{pennycook2019fighting}, even when those workers are informed that their ratings may influence social media feed rankings \cite{epstein2019letting}.
 
Despite these reasons for optimism, there is some debate about whether crowds will be able to produce helpful and informative annotations, especially when additional context or supporting evidence is required. For example, some suggest that highly informed crowdworkers --- even aided by algorithms --- still underperform relative to professional fact-checkers \cite{godel2021moderating}. Other studies show that a sufficiently large panel of informed crowdworkers can outperform a trio of expert journalists \cite{resnick2021informed}. One study suggests the mechanism by which professional fact-checkers outperform students and PhD historians in evaluating web information is a higher propensity for ``lateral reading,'' in which the evaluator leaves the primary source and checks the information against several other sources  \cite{wineburg2019lateral}. 
 
Beyond basic classifications of news accuracy, crowdworkers have been shown to capture different facets of misinformation. \citet{soprano2021many} 
found that with a small amount of instruction, crowds were able to effectively discern subtle characteristics like neutrality, comprehensibility, and completeness. This is useful since there are several operational definitions of misinformation within the academic literature \cite{ruths2019misinformation, ribeiro2019microtargeting}. 

These results suggest that annotations that provide outside sources and reflect consensus across the viewpoints of a balanced crowd could approximate the accuracy and efficacy of professional fact-checkers while achieving faster, broader results and not relying on a single source of truth.

We note that unlike the above research, our application does not involve crowdworkers producing annotations on specific posts. Rather, we invite people on social media to annotate posts of their choosing, and to indicate which annotations, on which posts, are the most helpful to them. 

Since Birdwatch is an open-source project, it has sparked at least a handful of studies published in the public domain. Some studies suggest that contributors to Birdwatch do not choose which content to rate at random, but instead tend to rate content written from an opposing political perspective. Similarly, they choose not to rate content written from a perspective consistent with their own \cite{allen2022birds}. Other studies look at more descriptive features of the community --- finding, for example, that Birdwatch notes are more likely to be written for misleading than non-misleading content, and that notes on posts from influential users with many followers are associated with lower consensus in note rating \cite{prollochs2022community}. Finally, some researchers analyzed an earlier, basic Birdwatch algorithm for evaluating note quality. This system used a supermajority algorithm where notes needed 84\% of their ratings to be ``helpful'' as opposed to ``unhelpful'' in order to be recommended by Birdwatch. Researchers identified, as one might expect, that such an algorithm could be vulnerable to coordinated abuse \cite{crisman2021identifying, mujumdar2021hawkeye}.  

The current work is different in important ways from prior Birdwatch studies. As of this writing, no publicly available studies tackle the problem of selecting annotations. Rather than descriptively analyzing the raw Birdwatch data, our work focuses on building an algorithm that selects notes that inform understanding for a diverse population of users better than the supermajority algorithm baseline. Our study is also not limited to analyzing those who contribute to Birdwatch itself. Instead, we study how annotations impact understanding among representative samples of Twitter users. 

\subsection{Bridging-based Ranking Algorithms}

Inspired by the finding that simply breaking echo chambers can actually increase rather than decrease polarization \cite{bail2018exposure}, there has been recent optimism about the potential of algorithms or platforms that prioritize and reward content that bridges polarized divides. This is in contrast to traditional engagement-based content-ranking mechanisms \cite{yu2021affective, stray2021designing}. 

Concurrent with the development and release of the algorithm described in this paper, a nascent field has coined the term ``bridging-based ranking'' to describe ranking mechanisms that reward behavior that ``leads to positive interactions across diverse audiences, even when the topic may be divisive'' \cite{recommendation2022bridging}. This white paper identifies the Birdwatch algorithm described in this work as an example. In our application, a bridging-based algorithm could capture points of agreement across otherwise polarized subgroups. It could help set better norms for crowd-based discussions of social media posts by rewarding content that is informative and appreciated by a wide range of people. While previous studies of polarized crowd work found lay people can struggle to generate accurate assessment of what is misleading, research on bridging-based algorithms offers a possible solution. 

As the idea is relatively new, there are a limited number of existing applications. One group identified self-reported Democrats and Republicans on Twitter and created a "bi-partisanship leaderboard" of users whose Tweets were liked by members of both parties \cite{bail2021breaking}. One participatory democracy tool, Polis, clusters the output of a matrix factorization to split users into opinion clusters, then elevates content with high group-informed consensus (agreement across the clusters) \cite{small2021polis}.

The Birdwatch bridging-based ranking algorithm we present in this paper uses matrix factorization as an unsupervised way to build representations of the (polarized) opinion space, but unlike previous approaches that require a clustering of users or items into distinct groups, Birdwatch's algorithm bridges across viewpoints in a continuous space, without the additional assumptions that clustering involves. Another novel contribution of our work is the demonstration, via survey data, that bridging-based scoring works: it selects notes that are seen as helpful by users with diverse self-reported political affiliations better than a baseline non-bridging algorithm.

\section{Method}

Before bridging-based algorithms can be applied, we need data that captures feedback from the crowd. We use data from Birdwatch, where contributors may add context to Tweets and rate the contributions of others. In this section, we detail how Birdwatch works, and the dataset upon which we developed our algorithms. We also discuss the collection and analysis of survey data used to understand the effect of the algorithm in the context of the Birdwatch ecosystem. 

\subsection{Dataset}

\begin{figure}
    \centering
    \includegraphics[width=0.9\columnwidth, trim={10px 0 0 0},clip]{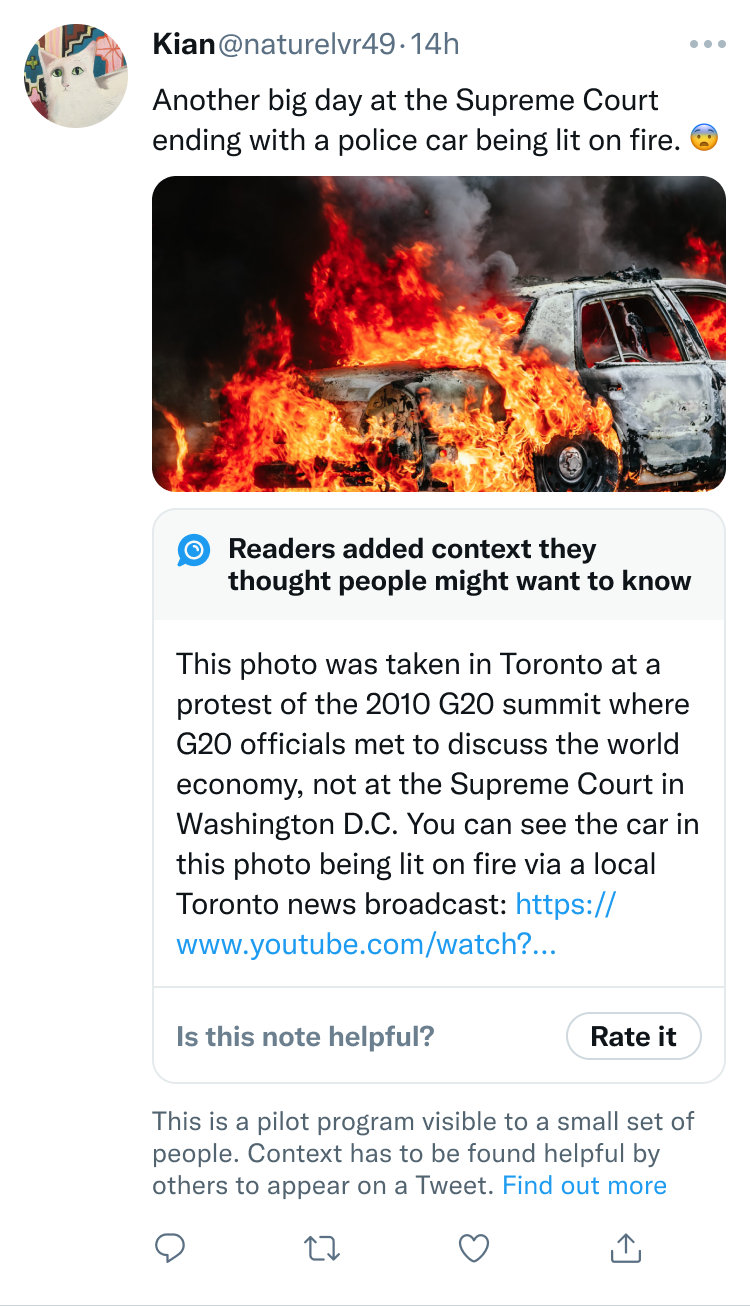}
    \caption{An example Tweet and note showing how a Tweet looks to a Birdwatch contributor (or crowd viewer) when the Tweet has a currently rated helpful note. The note is displayed as context immediately below the Tweet, with no extra clicks required to see it, wherever the Tweet is shown (e.g. on the home timeline, or after clicking on Tweet detail)}
    \label{fig:crh_note}
\end{figure}

The first set of users able to sign up to participate in the pilot of Birdwatch were Twitter users based in the United States who were able to provide a verified phone number from a trusted U.S.-based phone carrier, had 2-factor authentication enabled, and had not had a Twitter rule violation in the past year. The pilot began on Jan. 25, 2021\footnote{\url{https://twitter.com/TwitterSupport/status/1353766523664531459}}. 

Contributors are able to propose free-text notes to add any Tweets they view as potentially misleading and answer a small number of multiple choice responses about the associated Tweet as shown in Figure \ref{fig:note_form}. The response form asks the users questions about the Tweet, in particular asking whether the Tweet is potentially misleading and the primary reasons leading them to this conclusion.

Separately, contributors are prompted to rate the quality of notes made by other contributors as ``helpful'' or ``not helpful'' on an ongoing basis, and optionally provide multiple choice responses about why, shown in Figure \ref{fig:rating_form}. Later updates to the rating form added the option to rate notes ``somewhat helpful.''

Users can suggest notes on any Tweet they see while normally using the regular Twitter app, but they also have access to a separate Birdwatch home page (shown in Figure \ref{fig:bw_home}). The Birdwatch home page contains three tabs with timelines of Tweets: ``Needs Your Help,'' for Tweets that have notes with less than 5 ratings (the minimum number to receive a status label), ``New,'' a chronological timeline of Tweets with the most recently written notes, and ``Rated Helpful,'' where users can see all Tweets that recently received Birdwatch notes that received "Currently Rated Helpful" status based on their ratings by other users (the highest rated notes that would be shown to Twitter users in-app).

As users browse Twitter and the Birdwatch home page, they can see which Tweets have notes (since the notes are displayed just below the Tweet for Birdwatch pilot participants as in Figure \ref{fig:tweet_screenshot}). Despite the existence of the Birdwatch home page and Needs Your Help tab, a large majority of notes and ratings are made on Tweets and notes people found while organically using Twitter. As a consequence, users that follow a Tweet's author (or follow people who are likely to Retweet or Reply to a particular Tweet, etc.) are significantly more likely to write notes and rate notes on that Tweet than Tweets from other users.

After the initial environment had been available to contributors for 45 weeks, 15,368 notes had been generated on 10,321 Tweets by 2,558 users. 7,544 raters generated 93,560 helpful ratings, 56,328 not helpful ratings, and 10,789 somewhat helpful ratings, for an average of 58 percent helpful ratings.

\section{Algorithm}

 Here we detail the matrix factorization (MF) algorithm\footnote{In total, the authors evaluated several families of algorithms for aggregating the note ratings in order to label a note as helpful or not based on note-rater data from the initial rollout. The algorithm presented here performed best across our criteria of informed understanding and breadth of helpfulness}. The algorithm was trained from the note-rater matrix from the initial Birdwatch rollout. These data consist of a sparse matrix that encodes, for each note and rater, whether that rater found the note to be helpful or unhelpful. For each user $u$ and note $n$, the matrix has entry $r_{un} \in \{0,1, \texttt{null}\}$, where $0$ indicates an unhelpful rating, $1$ indicates a helpful rating, and \texttt{null} indicates the absence of a rating. 

The factorization of this sparse, high-dimensional matrix seeks low-dimensional latent vector representations of users and items that reconstruct the full ratings matrix with low error when multiplied. These factor vectors can be interpreted as explaining the affinity of certain users for certain items, as similar vectors multiplied together yield a higher dot product (rating), and vectors with opposing signs yield a lower dot product (rating) \cite{funk2006netflix, koren2009matrix}. In our application, this representation space identifies whether notes may appeal to raters with specific viewpoints. By controlling for this, we are able to identify notes that are more broadly appealing than might be expected given the viewpoint of the note and its raters.
In particular, we predict each rating as:

\[\hat{r}_{un} = \mu + i_u + i_n + f_u \cdot f_n\]

where the prediction is the dot product of the user and notes’ vectors $f_u$ and $f_n$ added to the sum of three intercept terms: $\mu$ is a global intercept term, accounting for an overall propensity of raters to rate notes ``helpful'' vs ``not helpful''; $i_u$ is the user’s intercept term, accounting for a user's leniency in rating notes ``helpful''; and $i_n$ is the note’s intercept term, accounting for idiosyncratic ``helpfulness'' of the note beyond that explained by rater viewpoints and leniency. This note intercept term, $i_n$ is used to assign a single global score to a note. We note that this represents a departure from traditional use of this algorithm to personalize content as in a recommender system.

Due to the extreme sparsity of the note-rating matrix (on average, each note receives just 10 ratings and each rater provides 21 ratings, with a median of 4 ratings per note, and 8 ratings per rater), estimation of model parameters is noisy and under-determined\footnote{The ratings filters were designed achieve a minimal density of the rating matrix, otherwise parameters remained unrestricted and in some cases impossible to estimate.}. To mitigate this challenge, we restrict to notes that have at least 5 ratings and raters who have completed at least 10 ratings. After this step, we have 7,088 notes and 124,241 ratings (70,783 helpful, 45,239 not helpful, 8,219 somewhat helpful) from 3,261 raters, with an average of 18 ratings per note (median=10) and 38 ratings per rater (median=22).

Further, in our task we particularly value precision (having a low number of false positives) over recall
(having a low number of false negatives) due to risks to our community and reputation from increasing visibility of low quality notes. With this in mind,
we use much higher regularization on the intercept terms, $\{\mu, i_u, i_n\}$.
This encourages a representation that uses user and note factor vectors to explain as much variation in the ratings as possible before fitting additional note- and user-specific intercepts. As a result, for a note to achieve a high intercept term, it must be rated helpful by raters with a diversity of factor vectors.

To estimate the model parameters, we minimize the following regularized least squared error loss function via gradient descent over the dataset of all observed ratings $r_{un}$:

\[\textrm{min}_{\{i,f,\mu\}}\sum_{r_{un}} (r_{un} - \hat{r}_{un})^2 + \lambda_i (i_u^2 + i_n^2 + \mu^2) + \lambda_f (||f_u||^2 + ||f_n||^2)\]

Where $\lambda_i$ (0.15), the regularization on the intercept terms, is currently 5 times higher than $\lambda_f$ (0.03), the regularization on the factors.

To avoid overfitting on our small dataset, we use one-dimensional factor vectors. Additional factors added little explanatory power \footnote{RMSE on held-out samples decreased from .076 to .073 when adding a second factor} and reduced interpretability and replicability. (Though we expect to expand dimensionality as the contributor base grows.) 

\begin{figure}
    \centering
    \includegraphics [width=1.05\columnwidth, trim={10px 0 0 0},clip]{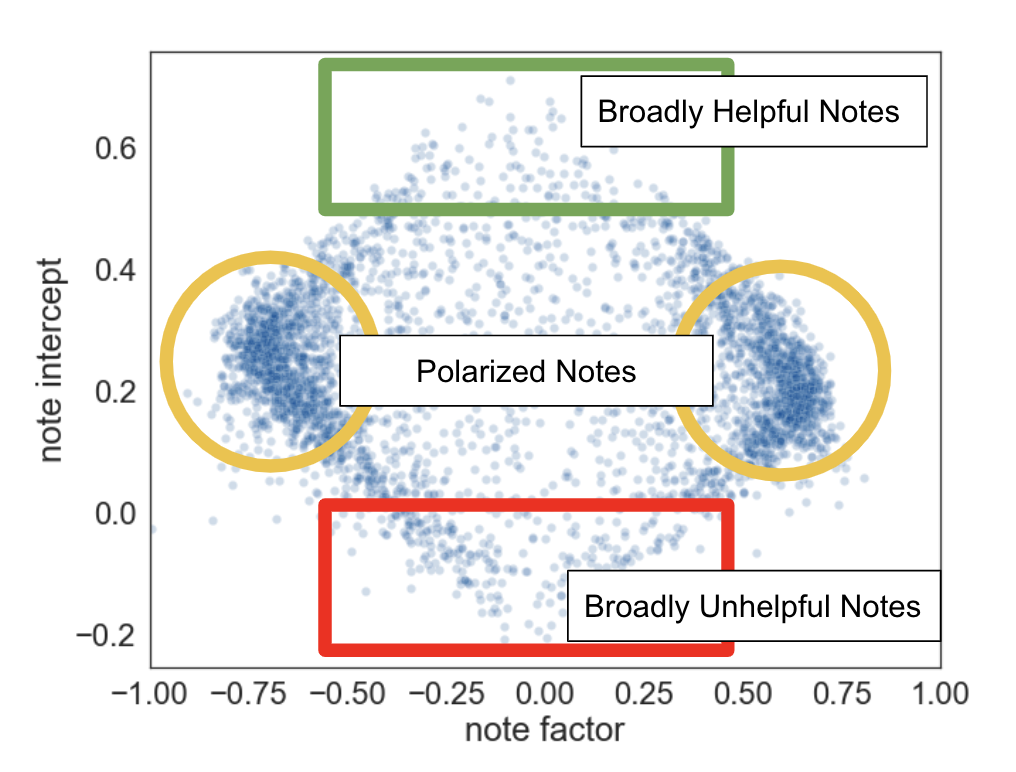}
    \caption{Distribution of estimated note intercepts (y-axis) and factors (x-axis) from the note selection algorithm.}
    \label{fig:diamond}
\end{figure}

The resulting note intercept and embedding scores can be seen in Figure \ref{fig:diamond}. Note intercepts are approximately Normally distributed, with most notes being highly polarized. Notes with the highest and lowest intercepts tend to have factors closer to zero.

We then apply the following thresholds to output note labels given note intercept scores: a note is ``currently rated helpful'' if its intercept is at least 0.40, ``currently rated not helpful'' if its intercept is at most -0.08, otherwise the note gets the ``needs more ratings'' label if its intercept is between -0.08 and 0.40.

\subsection{User Helpfulness Scores}

After first running this algorithm on all ratings that meet the minimum note and rating density requirement, we use the initial estimates of note intercepts to assign provisional ``helpful,'' ``not helpful,'' and ``needs more ratings'' labels using the thresholds described above. We then compute a set of user helpfulness scores using these provisional note labels in order to determine which raters should be filtered out before re-running the matrix factorization algorithm without them. The idea behind these helpfulness score filters is to prevent very low-quality users from influencing note scoring.

\subsubsection{Rater Helpfulness}

Rater Helpfulness is the proportion of a rater's ``valid ratings'' that match the note's provisional label.

In estimating ``rater helpfulness", we consider only ``valid ratings,'' which we define as ratings on notes provisionally labeled ``helpful'' and ``not helpful'' (not ``needs more ratings''), restricted to the first 5 ratings made within the first 48 hours of a note's existence. Because 5 ratings are required to receive a label\footnote{This criterion was later changed to include any rating before the first ``helpful''/``not helpful'' label was published}, this ensures that ratings were made before the note received a label. Because ratings data is published with a 48 hour lag, this helps ensure that the rating was made before any data was published regarding other ratings, reducing the possibility that raters can copy existing note labels or the ratings of others to obtain a high rater helpfulness score. Note that these rules only determine which ratings are used to determine a rater's rater helpfulness score, not when scoring notes.

To be included in the second and final helpfulness-filtered matrix factorization, raters must have made at least one ``valid rating'' and have a rater helpfulness score of at least 66\%, i.e. at least $66$\% of their valid ratings matched the note's provisional label.

\subsubsection{Author Helpfulness}

There is no requirement to have authored any notes in order for a user's ratings to be included in the final matrix factorization. However, if a user did author any notes that have received at least 5 ratings, that user must meet the following criteria for their ratings to be included:
\begin{itemize}
\item The number of provisionally ``helpful'' notes written must be at least 5 times the number of ``not helpful'' notes written
\item The average intercept of all notes they have written must be at least 0.05
\end{itemize}

These filters remove roughly half of both raters and ratings and results in final scores for 5,787 notes.

\subsection{Final Scoring}

The algorithm's final output is given by running the matrix factorization again on the filtered set of users, then applying the $\geq0.40$ and $\leq -0.08$ thresholds described above to produce the final note labels. In summary, the algorithm's steps include:
\begin{itemize}
    \item Filter out raters with <10 ratings or notes with <5 raters
    \item Fit matrix factorization and compute provisional note labels using the thresholds 
    \item Compute user helpfulness scores, then filter out all ratings from users that didn't meet the helpfulness score criteria
    \item Fit matrix factorization on the final filtered dataset, and compute final note labels using the thresholds

\end{itemize}

For evaluation, we compare notes selected by our algorithm in the April survey wave (described below) to an average over all notes and those selected by supermajority voting (raw rate of users voting ``helpful'' $\geq.84$). The former is meant to represent the null hypothesis that showing any annotation of reasonable quality would decrease belief in the annotated claim.

\section{Survey}

We relied on three waves of a survey experiment to evaluate the quality of our algorithm. The survey experiment showed all respondents Tweets that had been annotated on Birdwatch, and thus all Tweets contained claims that were potentially misleading (see example in \ref{fig:crh_note}). To examine the effect of annotations, a random subset of respondents was also shown the annotation. All participants were ultimately debriefed about the content they saw at the conclusion of the survey. 

Each wave drew on a stratified random sample of Twitter users and excluded those already participating in the Birdwatch program (sampling strata included age, gender, and user activity). We fielded an initial pilot wave in August 2021, but focus in this study on Wave Two and Wave Three. Wave Two, used in algorithm selection, entered the field in December 2021 (N=7,387). Wave Three, used to evaluate the algorithm selected for production use, entered the field in April 2022 (N=15,935). All respondents were recruited on the Twitter platform. 

In these surveys, a subset of respondents were shown Tweets alone (\emph{control} group), another subset was shown Tweets plus annotations (\emph{treatment} group), and a third subset was shown Tweets plus annotations but a different questionnaire (\emph{evaluation} group). Each respondent in the \emph{treatment} and \emph{control} groups was asked whether they thought the Tweet they saw was accurate, whether they agreed with a statement summarizing the main claim of the Tweet, and whether they agreed with a statement summarizing the main claim of the note. Each respondent in the \emph{evaluation} group was asked whether they found the annotation to be helpful or unhelpful for understanding the issue discussed in the Tweet. Each of these key dependent variables is measured on a five-item ordinal scale, plus a ``don't know'' option. Respondents were also free to skip any question.

After the condition-specific modules of the questionnaire, all respondents were then asked to optionally self-report age, gender, race, and party identification. Tables \ref{tbl:descriptives_w2} and \ref{tbl:descriptives_w3} report both unweighted and weighted sample statistics. Weights are derived with the variables used in sampling. They adjust both for sampling variability and for nonresponse, and extreme weights have been trimmed to reduce the extent to which the variance of estimates is inflated due to unequal weighting. \cite{biemer2008weighting}

To answer \textbf{RQ1}, we evaluate ability to inform understanding by comparing agreement with the main claim of the Tweet between participants assigned to the \emph{treatment} and \emph{control} groups. Responses from the \emph{evaluation} group regarding the subjective helpfulness of the annotations are examined by self-reported demographic features to evaluate broad appeal. To answer \textbf{RQ2}, we examine the relationship between two measures: perceived Tweet accuracy and Tweet claim agreement. We examine whether appraisals of Tweet accuracy can act as a proxy for Tweet agreement. We examine the effect of annotations on responses using each measure separately, and we also explore whether the association between the two variables is different in treatment and control groups. 

\textbf{Considerations for Human Subject Data:} Prior to being asked for their consent to participate in this study, respondents were informed that the research was about conversations on Twitter concerning news, political events, and social events, and were informed of the approximate time required to complete the study. All questions were voluntary, and the participants were informed they could stop at any time. In addition, participants were provided instructions for how to post-hoc opt-out of the survey if they so wished by contacting the researchers. Data access was limited to a small group of researchers.

When asking participants to review Tweets that potentially contained misinformation, we took care to choose Tweets that would not be harmful if they were shown to users. To be included, Tweets could not be in violation of any Twitter policies, and could not use forms of persuasion that might cause offline or online harm, such as invoking images of children while advocating against the COVID vaccine \footnote{These criteria were determined by the authors in consultation with experts in Internet Safety}. In Wave 2 and Wave 3, participants who completed the survey were additionally debriefed about Birdwatch as the source of the content they saw. To ensure that participants were not deprived of helpful context, upon completing the survey, we showed participants in the control group the same note that was shown to those in the treatment group.

\subsection{Algorithm Evaluation}

As we state at the outset, one of the two design goals of our system is to identify notes that \emph{inform users' understanding} about the issues raised in Tweets. When users who see an annotation are less likely to agree with the substance of a potentially misleading Tweet than those who do not, we say that note is able to \emph{inform understanding}. 

We sought to measure whether the algorithm selected notes that, on average, could inform users' understanding about issues raised in Tweets. For inclusion in each survey wave, we selected sets of contributor-generated annotations on contributor-selected Tweets to emulate, as closely as possible, the full set of annotations that might appear on Twitter during the period immediately before each survey entered the field.

For the December 2021 wave, we began with 75 notes created during a two-week period prior to the survey entering the field that were estimated to be among the top 20\% of all notes by quality according to any one of eight different note-ranking algorithms (see Appendix Section \ref{note_selection}). We then ruled out notes according to a set of pre-specified criteria concerning the generalizability of results (such as notes that do not assert that the related Tweet is misleading, notes on deleted Tweets, or notes on Tweets that only make sense in the context of a Twitter thread), Twitter policy, and ethical concerns (details in Appendix Section \ref{note_selection}). This resulted in a final set of 38 notes included in that survey. 

In the April 2022 survey wave, we began with 103 notes created during a four-week period prior to the survey entering the field, and with an algorithm score at or above a threshold that subjectively appeared to be an appropriate minimum bar for note quality, but that was below the threshold at which a note would be ``rated Helpful'' and shown on Twitter to people outside the Birdwatch pilot. We then followed the same procedure, arriving at a final set of 46 notes. We selected a longer window for inclusion in the algorithm evaluation wave in order to arrive at roughly the same number of notes as in the algorithm selection wave, despite only having one set of notes to draw from rather than the eight sets included previously.

To estimate the overall effect of annotations on users' understanding of issues raised in Tweets (\textbf{RQ1}), we compared rates of agreement between respondents who were shown notes to those who were not (denoted ``Any'' in all tables). To evaluate the specific effect of the supermajority and the algorithm, we do the same comparison while subsetting to notes that received supermajority votes or that the algorithm scored highly, respectively\footnote{Note that in Wave 3 we only include notes scored highly by the algorithm.}. We collapsed the original ordinal response scales into binary measures that take the value 1 if a respondent indicates they ``agree'' or ``somewhat agree'' with a statement, and 0 otherwise. We then estimated intent-to-treat effects for Birdwatch notes --- that is, the effect of being \emph{assigned} to view a note, an effect we expect to be attenuated by the one-way nonresponse of participants who do not read notes they are assigned to view --- using simple differences in proportions between users included in the treatment and control groups. In the case of Wave 2, these calculations are subset by algorithm. Finally, we used binomial logit models to explore whether estimates of population averages obscured substantively important differences between subpopulations, either correlated with respondent attributes or explainable by heterogeneous treatment effects. 

To explore heterogeneous treatment effects by party ID, we fit binomial logit models of the form: \\

\noindent
$ log \frac{p}{1-p} = \beta_0 + \beta_1^{*} \cdot \text{Treatment} + \beta^{*}_2 \cdot \text{Party ID} + \beta^{*}_3 \cdot \text{Age} + \beta^{*}_4 \cdot \text{Gender} + \beta^{*}_5 \cdot \text{Race} + \alpha^{*}_1 \cdot \text{Treatment} \cdot \text{Party ID}$ \\

\noindent where variables superscripted with an asterisk are categorical, and $\beta^{*}_2$, for instance, is here shorthand referring to a set of coefficients, one for each of a series of mutually exclusive and exhaustive dummy variables (see Tables \ref{tbl:w2_tw_logit_models}, \ref{tbl:w2_nt_logit_models},  \ref{tbl:w3_logit_models}). In order to illustrate these effects with simple graphical displays, we additionally fit binomial logit models including only assignment to treatment, party ID, and an interaction term\footnote{An initial pilot study found that while there are substantively important differences between different genders, ages, and races in the baseline propensity to believe misinformation, the strongest and most substantively significant differences in the actual effect of exposure to annotation had to do with ideology and/or partisan attachment. In that pilot survey, we found (through a likelihood ratio test) that self-reported partisan attachment had greater explanatory power than self-reported ideological placement; because the two are difficult to disambiguate, we report models including party ID only.}. These are used to generate displays showing expected probabilities for both treatment and control by party ID, which is the primary heterogeneous effect of interest.

Next, we evaluate whether notes are not only \emph{informative}, but also \emph{perceived as helpful}. Using measures included in the survey, we analyze if Birdwatch contributors and our rating algorithms identify sets of notes that have broad appeal to Twitter users (\textbf{RQ1}).

Finally, we compare these findings in light of conventional measures of perceived content accuracy, and evaluate how these measures perform relative to informed understanding (\textbf{RQ2}).

\section{Results}

Upon evaluating the MF algorithm, we found positive signals for \textbf{RQ1} and \textbf{RQ2}. First, the algorithm shows a strong and statistically significant effect on helping inform understanding about potentially misleading information (\textbf{RQ1}). When participants in the treatment condition were shown the most helpful notes according to the algorithm, they were less likely to be agree with the content of the Tweet compared to the control group. Second, a wide majority of all users (64\% in Wave 3) find notes selected by the algorithm to be subjectively helpful (\textbf{RQ1}). Third, we find that perceived accuracy of a Tweet and agreement with the core claim of a note are only moderately correlated, and that this correlation is lower among those who see notes as well as Tweets than it is among those who see Tweets alone (\textbf{RQ2}). 

\subsection{RQ1}
On average, according to Wave 3 results, notes selected by the algorithm reduce the likelihood of agreeing with the substance of a potentially misleading Tweet by about 26\% among Twitter users (weighted binomial logit, p << 0.001, N=6,046, table not shown). Similarly, users who see a note are more likely to agree with the substance of that note. 

\begin{figure}
    \centering
    \includegraphics [width=1.0\columnwidth]{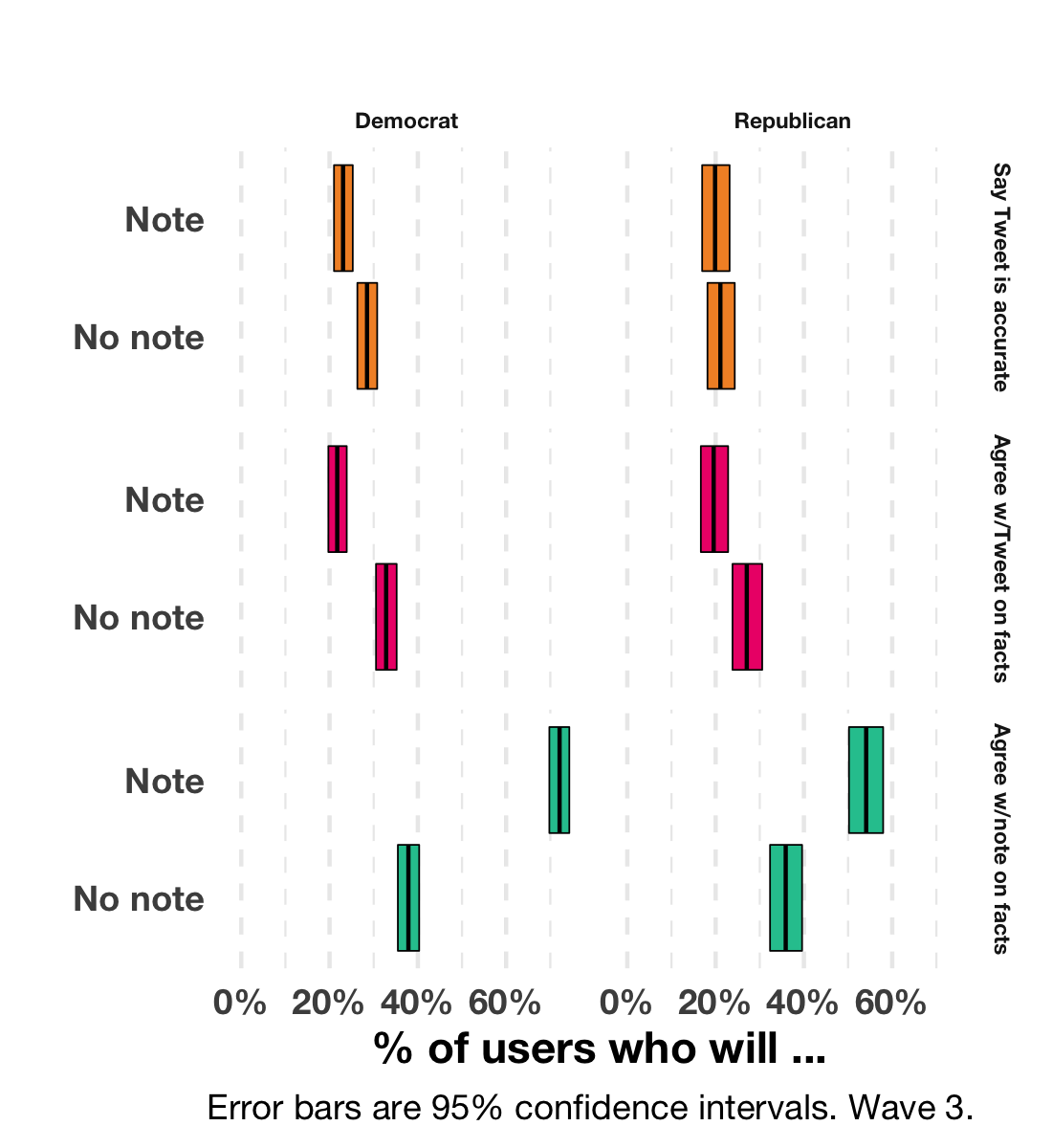}
    \caption{Predicted Tweet agreement in survey for note and no note conditions by self-reported political party.}
    \label{fig:w3_belief_results}
\end{figure}

We find no statistically significant difference in the effect of notes for Democrats and Republicans (Table \ref{tbl:w3_logit_models}; results from model 1 in this table are illustrated in Figure \ref{fig:w3_belief_results}).

Notes selected by the algorithm also reduce uncertainty among survey respondents. Relative to respondents who do not see a Birdwatch note, users who see a note are 36\% less likely to respond that they are ``not sure'' when asked if they agree or disagree with the substance of that Tweet. Further, they are about 52\% more likely to say that they strongly disagree (Fig. \ref{fig:w3_tw_responses}).

Similarly, 29\% of people who do not see a note would say they strongly disagree with the claims in a Tweet --- compared to 44\% of those who do, a 52\% increase.

\begin{figure}
    \centering
    \includegraphics [width=1.0\columnwidth]{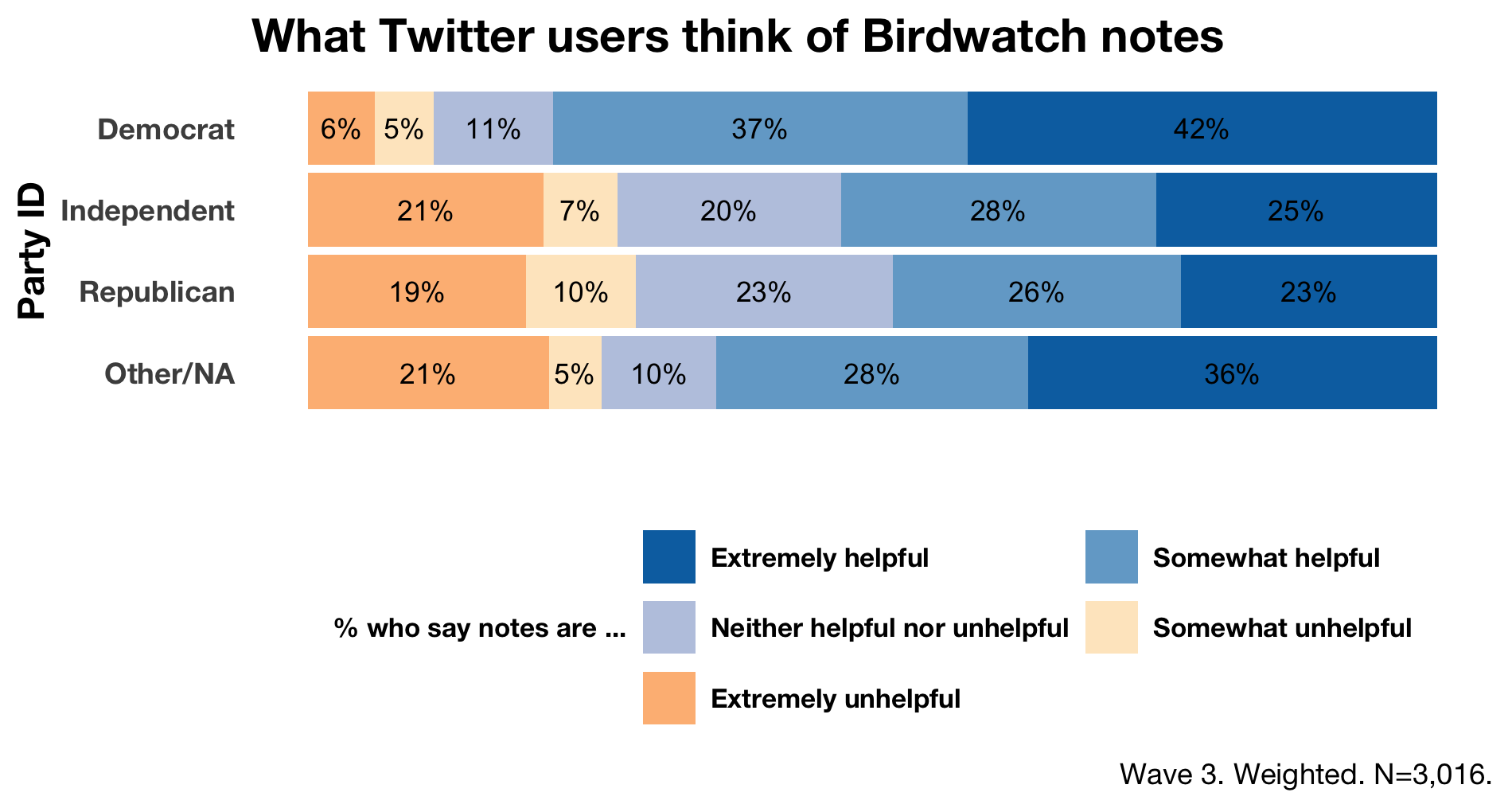}
    \caption{Reported helpfulness of Birdwatch notes, by party affiliation.}
    \label{fig:w3_helpful_by_pid_long}
\end{figure}

While other research shows that people are divided on whether social media companies should label potentially misleading content in general \cite{mitchell2021more}, they are far less divided where Birdwatch is concerned. Nearly 80\% of all users either find chosen notes helpful or find them neither helpful nor unhelpful --- for adherents of either major political party, this number is 70\% or higher (Fig. \ref{fig:w3_helpful_by_pid_long}). Additionally, nearly 80\% of self-identified Democrats and close to half of self-identified Republicans find the notes selected by the matrix factorization algorithm to be somewhat or extremely helpful for understanding the issues raised in Tweets (Table \ref{tbl:helpful_bypid_w3}).

\begin{figure}
    \centering
    \includegraphics [width=1.0\columnwidth]{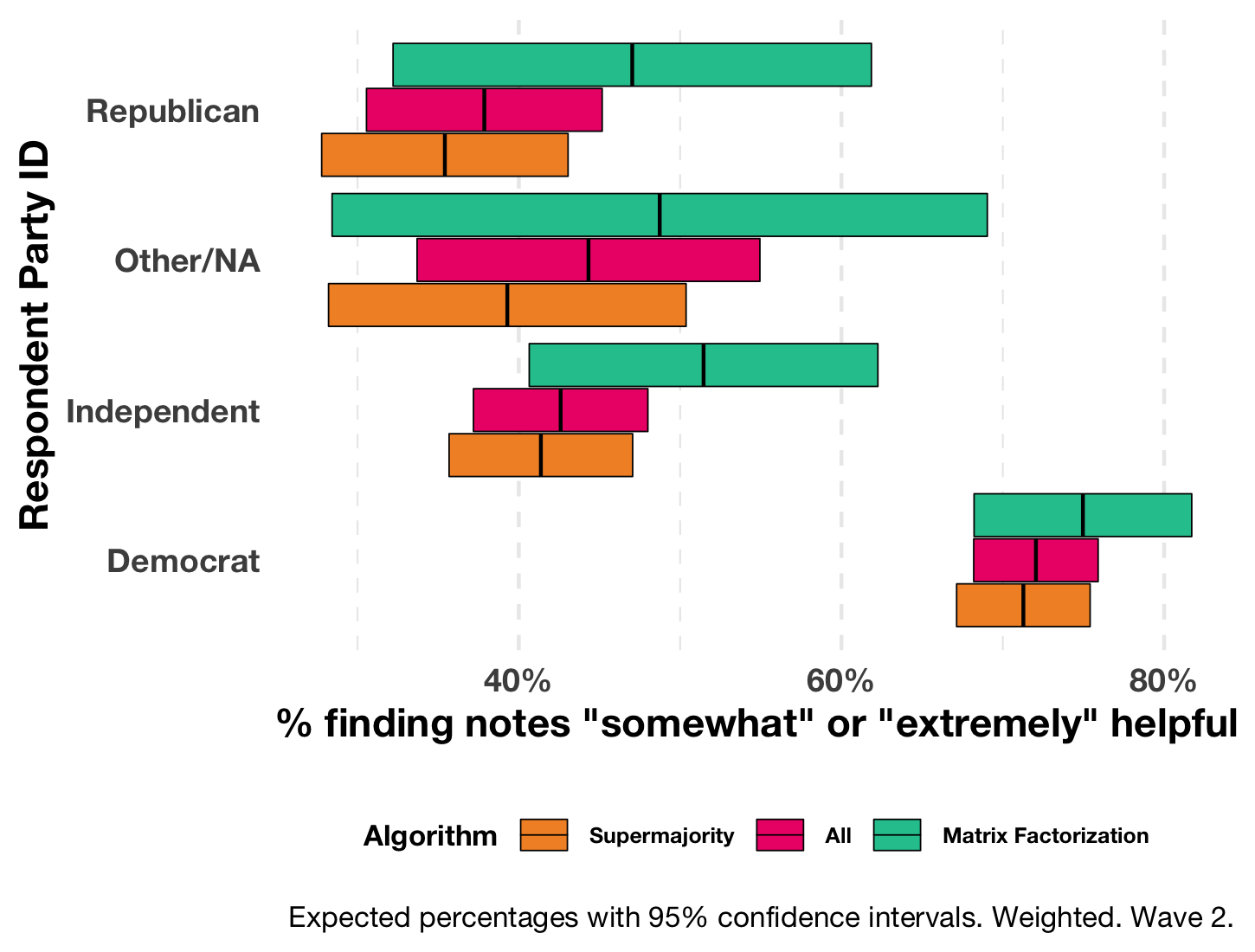}
    \caption{Perceived helpfulness of notes, by algorithm, by PID, Wave 2.}
    \label{fig:w2_helpful_bypid_byalgo}
\end{figure}

These results are an improvement upon the performance of simple supermajority voting by contributors. In Wave 2 of our survey, we tested matrix factorization approaches against the aforementioned supermajority-voting-based algorithm and found that matrix factorization algorithms, by accounting for a diversity of perspectives, select notes that appear to be more informative and are seen as more helpful by a more politically diverse majority of people on social media (Fig. \ref{fig:w2_helpful_bypid_byalgo}).

People who saw any Birdwatch note selected by any of the algorithms we tested were, on average, 22\% less likely to agree with the substance of a misleading Tweet (weighted binomial logit, p $<$ .01). For notes selected by the original supermajority vote algorithm, the effect was identical (weighted binomial logit, p $<$ .01). In Wave 2, we found that Twitter users who saw notes chosen by the matrix-factorization algorithm were 41\% less likely to agree with the substance of a misleading Tweet than users who did not (weighted binomial logit, p $<$ .01). After controlling for age, gender, and party ID, the unweighted results (reported in Table \ref{tbl:w2_tw_logit_models}) are similar. 

The gap between our algorithm and baseline algorithms was more pronounced when considering perceived helpfulness. On average, notes chosen by the MF algorithm were perceived as about three percentage points more helpful compared to baselines by Democrats, but about 10 percentage points more helpful by Republicans (Fig. \ref{fig:w2_helpful_bypid_byalgo}).

\subsection{RQ2}

\begin{figure}
    \centering
    \includegraphics [width=1.0\columnwidth]{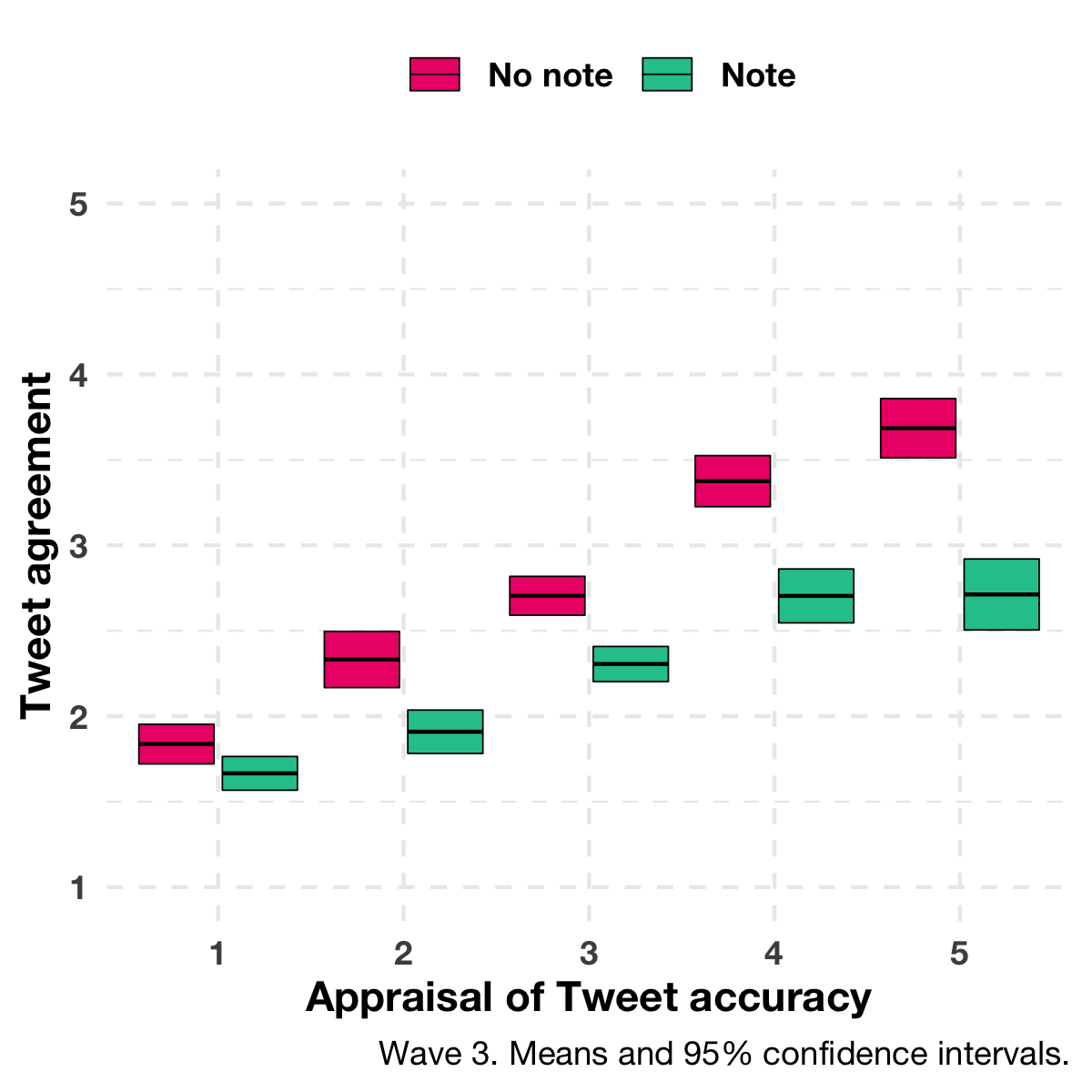}
    \caption{Bivariate association between agreement with substance of Tweet and appraisal of Tweet accuracy.}
    \label{fig:w3_accuracy_vs_belief}
\end{figure}

Our research also casts light on measurement differences between perceived accuracy and claim agreement in misinformation research. We present respondents in treatment and control conditions with three questions in series: First, concerning the accuracy of the Tweet; next, concerning agreement with a summary of the Tweet; and finally, concerning agreement with a summary of the note. Figure \ref{fig:w3_accuracy_vs_belief} illustrates the bivariate association between agreement with the Tweet and appraisals of accuracy in Wave 3 of the survey. As presented, both are on five-item scales where 3 is neither accurate/agree nor inaccurate/disagree, and larger numbers indicate more accuracy/higher agreement. 

In the control group, there is a modest correlation between appraisal of Tweet accuracy and agreement with the claims of the Tweet ($r=0.45$). Exposure to annotations attenuates this correlation ($r=0.29$). 

We find that measures of perceived accuracy are sufficiently noisy that an analyst who uses them may miss sizable effects (Fig. \ref{fig:w3_belief_results}). In particular, if we used appraisals of Tweet accuracy alone to evaluate whether additional context makes it less likely that people on social media will believe misleading information, we would fail to find a substantial or statistically significant effect (Appendix \ref{appendix_tables} Table \ref{tbl:w3_logit_acc}).

This finding relates to \textbf{RQ2} in that we suspect that studies that rely on proxy measures alone may not perform well in predicting informed understanding. The takeaway is that in the context of social media, it is important to differentiate between measuring the perceived accuracy of a misleading post and the likelihood of agreeing with misleading claims in a post. We reason that in the absence of corrective information, a misleading Tweet may appear to some people as being ``accurate'' as in factually accurate, and also ``accurate'' in the sense of speaking to some larger truth. In the presence of corrective information, many people will agree that a Tweet's specific \emph{factual} claim is not \emph{correct}, but continue to say that the Tweet itself is still \emph{accurate} --- because, we speculate, they feel the Tweet speaks to a larger truth.

\subsection{RQ3}

\begin{figure}
    \centering
    \includegraphics[width=1.0\columnwidth, trim={10px 0 0 0},clip]{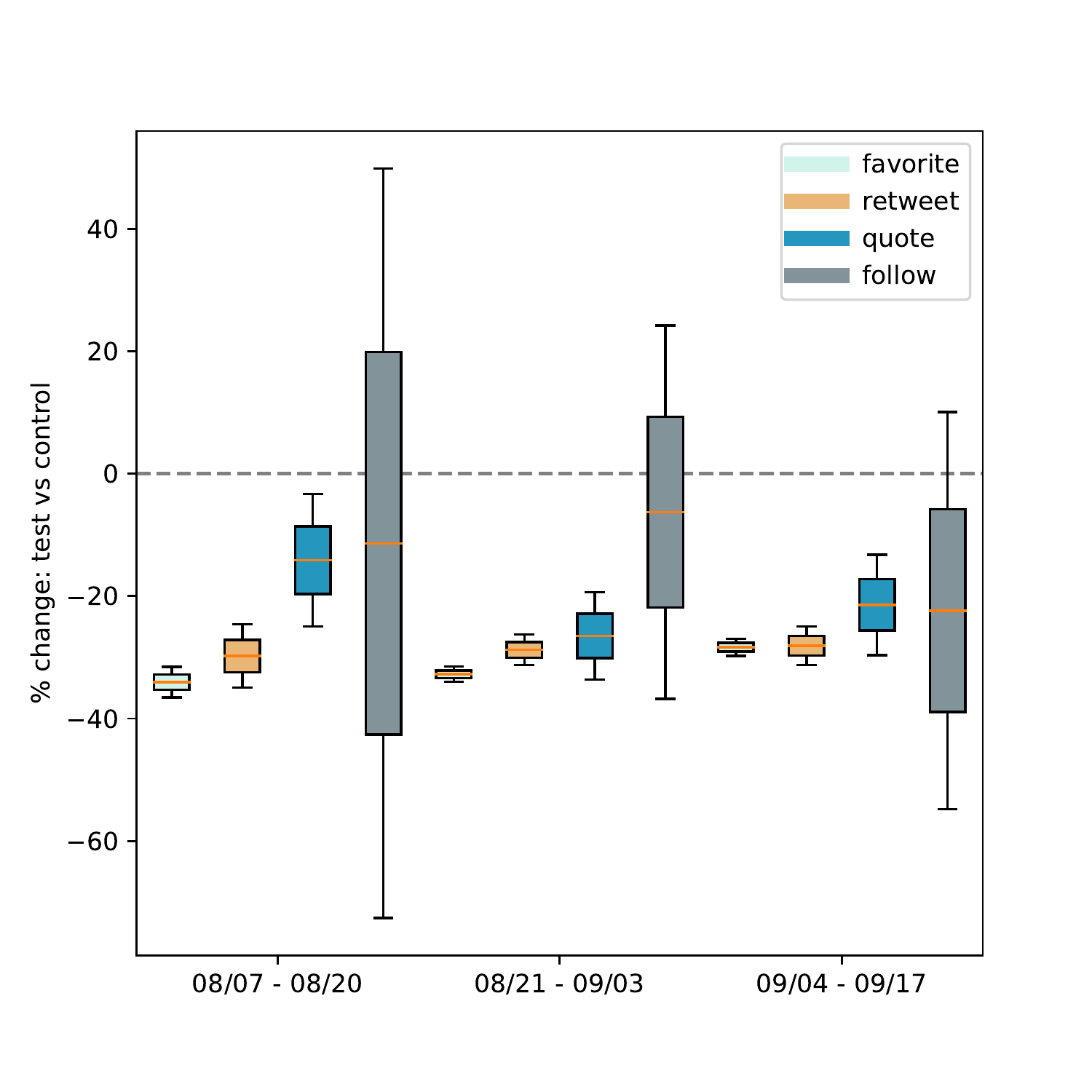}
    \caption{A/B test effects on engagement rates (interaction/impression). Error bars correspond to 95\% CI.}
    \label{fig:exp_results}
\end{figure}

As a final data point, A/B experiments conducted following the survey experiment compared engagement rates across users who were randomly assigned to be shown Tweet annotations compared to those shown no annotations. These annotations were presented (or not) during users’ normal use of Twitter. We found that those exposed to annotations on Tweets were 25-34\% less likely to like or retweet them compared to the control group (see Fig. \ref{fig:exp_results}). This finding likely reflects an underestimate due to limitations in data collection. Processing latencies in our data pipeline made it challenging to strictly identify users who had been treated - resulting in small rates of false inclusion and exclusion. These issues result in treatment dilution whereby we expect the measured experimental effects to be biased toward zero. 

\section{Discussion and Limitations}
This work represents an initial effort to confront the challenges of misinformation on social media beyond the laboratory and under realistic conditions. We sought to design an algorithm that can discern high-quality, user-generated notes that inform understanding and are found helpful by a wide range of people. 

We demonstrate a successful hybrid approach that draws on both crowd and algorithmic judgment to curate crowd-generated contributions. We demonstrate that by using a bridging-based ranking approach we are able to algorithmically select annotations that show strong evidence of causally informing users' understanding of issues raised in Tweets, regardless of their political attachments, while maintaining broad appeal. We additionally show that perceived accuracy is a noisy predictor of informed understanding, underscoring the need for research in this area to more directly measure the human-centered outcome of interest, rather than relying on proxy measures. 

This solution is open source and offers the public a transparent view onto how annotations are surfaced. The code, its history, and crowd-generated annotations are made available to the public for exploration or interrogation. 

There are limitations to the conclusions that may be drawn from this work. While we showed that certain notes can aid individuals across the political spectrum in understanding the subject matter of Tweets, it would be presumptuous to extrapolate that the effect of adding notes will have the same effect universally. 
In particular, if our survey participants are not the most informed or interested in the underlying issues to begin with, then the effects might be larger than they otherwise would. In other words, individuals who have stronger preexisting opinions might be harder to affect. On the other hand, if our survey participants are highly informed and interested in the underlying issues, then the effects might be smaller than they otherwise would.

We endeavored to ensure that our claims about the informativeness of annotations would be robust to heterogeneous treatment effects. We show that, on average, notes are informative for people of opposing partisan attachments. While we can rule out this category of heterogeneous effect, this is just one of a number of factors that might cause notes to be more useful for some people than for others. Similarly, it is unclear whether or how how our results will generalize outside of the context of the United States.

Birdwatch continues to be a rapidly evolving project, and here we note several challenges that are part of our ongoing work to improve the Birdwatch system.

Birdwatch notes are crowdsourced, and thus our study draws on a much more diverse pool of fact-check language compared to prior work. On the one hand, it provides an array of framing and perspectives around similar issues --- allowing the algorithm to select the least polarizing among notes of otherwise equal quality. On the other hand, notes are not guaranteed to meet a minimum bar of accuracy, and we must carefully design the system to avoid notes that are not in the spirit of improving understanding of issues. 

Again, because Birdwatch relies on crowdsourced data and a transparent algorithm, it is also important to ensure that the system is robust to adversarial manipulation. Creating an algorithm that is resistant to the actions of malicious actors who seek to undermine the project, to upvote certain groups, or to manipulate the ratings of particular groups deserves careful consideration. While such analysis is beyond the scope of this paper, our algorithm does provide some inherent resistance to certain forms of manipulation because of its bridging properties, and because it takes rater helpfulness into consideration before final note scoring.

The Birdwatch ecosystem may evolve to a different steady-state as community norms are generated, users gain knowledge, and the crowd becomes increasingly large. There are many possible outcomes. One is that the act of ranking combined with sui generis social norms creates incentives for annotation accuracy. In other possibilities, new areas for polarization may develop, and the algorithm may need to capture increasing diversity of perspective where none previously existed. 

Contributors decide which Tweets need additional context, and which notes are worthy of their rating and consideration. This has the potential to generate bias in the types of Tweets that get annotated. Initial evidence points to Tweet views as the primary predictor of Tweet annotation. As a Tweet gets more views, there is a marginally higher likelihood that a contributor will add context. As a consequence, it is likely that Tweets about high-salience issues (e.g. politics) will be more likely to be annotated compared to Tweets about low-salience issues. 

Another potential limitation is that bridging-based algorithms rely on some level of consensus across polarized groups --- that there is at least a small number of individuals from each side who can agree. By construction, the algorithm needs some cross-partisan agreement to function. It is possible this could lead us to be unable to surface high-quality notes for contentious political topics where a high degree of view polarization is expected. In an extreme case, this could lead the system to be effective at discerning quality notes on trivial topics. In practice, we have observed that Birdwatch produces high quality notes on contentious political and health topics, however this will be an important outcome to continue monitor as the program expands \cite{coleman2022helpful}.

A final consideration is how this method scales to modern social media settings. The rater-note matrix is bound to grow by orders of magnitude as more and more contributors are on-boarded to the system. This presents problems around how to compute the matrix factorization on a regular cadence to update note scores. The nuances of this issues are beyond the scope of this paper, but in short this issue is another where the system will continue to evolve.

\section{Ethical Considerations}

Here we describe the ethical considerations most salient to us as we conducted this research.

At the broadest level, we generate an algorithm that selects notes that have a high likelihood of informing understanding of the public on complex issues. In doing so, we incur a responsibility to understand how the algorithm works and where it might fall into trouble. To undertake proper vetting of this algorithm, we subjected it to rigorous tests that ensure it provides consistent results. That means actually reading the notes the algorithm selects (and those it does not select) to see if the choices made are sensible. We additionally used out-of-sample tests via the survey experiment to ensure our algorithm was not fitting to noise. The algorithm was trained prior to evaluation on the survey and analyzed by a researcher who did not develop the algorithm directly. It is also important to be transparent about the development and application of algorithms which will take on such a job, hence this paper.

\subsection{Acknowledgements}
    We thank Sarah Anoke, Lucas Neumann, Maria Antoniak, Maria Gorinova, Deblina Mukherjee, Emily Thai, Colin Fraser, Ryan Gallagher, and the entire Birdwatch engineering team for their contributions. We are also grateful to our advisory board (Amy X. Zhang, Adam Berinsky, David Rand, Sarita Schoenebeck, Jacob N. Shapiro), Christopher Bail, and advisors from the University of Chicago Center for RISC for their advice and feedback on earlier iterations of this project.

\bibliographystyle{ACM-Reference-Format}
\bibliography{birdwatch}

\clearpage

\onecolumn
\appendix

\section{Appendix: Link to Public Birdwatch Guide}

Additional information regarding the Birdwatch product can be found at \url{https://twitter.github.io/birdwatch}. The Birdwatch Guide can be found by following the above link, and contains current guidelines and rules for participating in Birdwatch. Note that Birdwatch is an evolving product, and some information may have changed since the writing of this paper.

\section{Appendix: Supplemental Survey Results Tables}
\label{appendix_tables}

The survey results presented in this paper rely on two survey waves (Waves 2 and 3) which were conducted on the Twitter platform in late 2021 and early 2022. In this section we display supplemental tables and figures showing basic demographic breakdowns, response proportions, and statistical models that we did not include in text for space considerations. 

We display the descriptive sample statistics for Waves 2 and 3 in Tables \ref{tbl:descriptives_w2} and \ref{tbl:descriptives_w3}. In Table \ref{tbl:helpfulness_w3}, we present the percentage of Twitter users who would agree with different assessments of the helpfulness of Birdwatch notes. The table includes survey-weighted estimates calculated using separate logit models for each valid response, along with the upper and lower bounds of 95\% confidence intervals. Unweighted results are substantially similar.

Table \ref{tbl:helpful_bypid_w2} reports the percentage of Twitter users finding Birdwatch notes at least "somewhat helpful" in Wave 2. Table \ref{tbl:helpful_bypid_w3} reports what percentage of Twitter users would find Birdwatch notes selected by the revised production algorithm, at the current threshold for note scores generated by that algorithm, to be at least "somewhat helpful," by party ID, and weighted by age and user state. The table includes estimates of population percentages derived using a survey-weighted logit model, along with the upper and lower bounds of 95\% confidence intervals. (Unweighted results are substantially similar.)

The remaining tables in this section display regression results from a number of models described in the paper. These include binomial logit models estimating the probability of agreeing with the substance of a note or the substance of a Tweet (Tables \ref{tbl:w2_tw_logit_models}, \ref{tbl:w2_nt_logit_models}, \ref{tbl:w3_logit_models}), and appraisals of Tweet accuracy (Table \ref{tbl:w3_logit_acc}). Table \ref{tbl:w3_tw_5pt_model} reproduces \ref{tbl:w3_logit_models} on the five-point scale used to collect the data, which allows for a direct comparison against unstandardized coefficients reported in the literature on journalistic fact-checks \cite{porter2021global}.

\begin{figure}
    \centering
    \includegraphics [width=1.0\columnwidth]{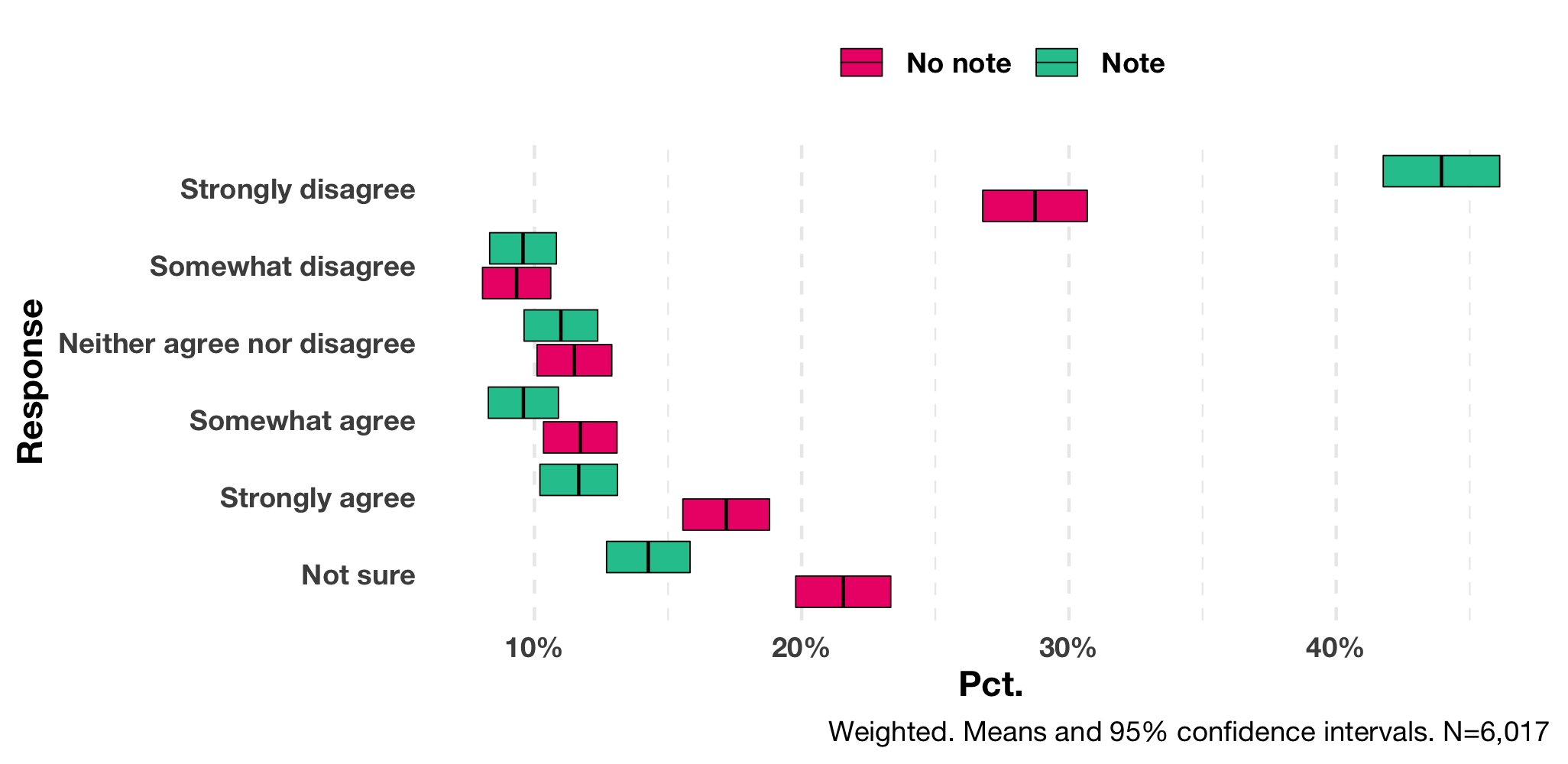}
    \caption{Weighted responses to ``Tweet agreement'' question in treatment and control groups, including ``not sure'' responses.}
    \label{fig:w3_tw_responses}
\end{figure}

Finally, Figure \ref{fig:w3_tw_responses} shows the weighted responses to the survey question asking whether respondents agree with the substantive claim in a Tweet.  

\begin{table*}[ht]

\caption{Descriptive sample statistics, Wave 2 (N=6,775)}
\centering
\begin{tabular}[t]{l|r|r|r}
\hline
Variable & N & Prop. (unwt) & Prop (wt)\\
\hline
Self-reported age: 18-24 & 1364 & 0.20 & 0.25\\
\hline
Self-reported age: 25-34 & 1509 & 0.22 & 0.23\\
\hline
Self-reported age: 35-44 & 1327 & 0.20 & 0.18\\
\hline
Self-reported age: 45-54 & 1188 & 0.18 & 0.16\\
\hline
Self-reported age: 55-64 & 900 & 0.13 & 0.12\\
\hline
Self-reported age: 65+ & 487 & 0.07 & 0.06\\
\hline
Self-reported gender: Man & 4056 & 0.60 & 0.59\\
\hline
Self-reported gender: Non-binary & 351 & 0.05 & 0.05\\
\hline
Self-reported gender: Other & 118 & 0.02 & 0.02\\
\hline
Self-reported gender: Woman & 2250 & 0.33 & 0.34\\
\hline
Self-reported race: Asian and no other & 284 & 0.04 & 0.05\\
\hline
Self-reported race: Black and no other & 501 & 0.07 & 0.07\\
\hline
Self-reported race: Decline to state & 402 & 0.06 & 0.06\\
\hline
Self-reported race: Hispanic (any other) & 834 & 0.12 & 0.13\\
\hline
Self-reported race: Other/multi-racial & 576 & 0.09 & 0.09\\
\hline
Self-reported race: White and no other & 4178 & 0.62 & 0.59\\
\hline
Self-reported party ID: Democrat & 3487 & 0.51 & 0.48\\
\hline
Self-reported party ID: Independent & 1786 & 0.26 & 0.26\\
\hline
Self-reported party ID: Other/NA & 423 & 0.06 & 0.07\\
\hline
Self-reported party ID: Republican & 1079 & 0.16 & 0.19\\
\hline
\end{tabular}
\label{tbl:descriptives_w2}
\end{table*}

\begin{table*}

\caption{Descriptive sample statistics, Wave 3 (N=9,108)}
\centering
\begin{tabular}[t]{l|r|r|r}
\hline
Variable & N & Prop. (unwt) & Prop (wt)\\
\hline
Self-reported age: 18-29 & 2493 & 0.27 & 0.34\\
\hline
Self-reported age: 30-39 & 1954 & 0.21 & 0.20\\
\hline
Self-reported age: 40-49 & 1910 & 0.21 & 0.19\\
\hline
Self-reported age: 50-59 & 1569 & 0.17 & 0.16\\
\hline
Self-reported age: 60+ & 1182 & 0.13 & 0.12\\
\hline
Self-reported gender: Man & 5766 & 0.63 & 0.64\\
\hline
Self-reported gender: Non-binary & 343 & 0.04 & 0.04\\
\hline
Self-reported gender: Other & 151 & 0.02 & 0.02\\
\hline
Self-reported gender: Woman & 2848 & 0.31 & 0.30\\
\hline
Self-reported race: Asian and no other & 320 & 0.04 & 0.04\\
\hline
Self-reported race: Black and no other & 612 & 0.07 & 0.06\\
\hline
Self-reported race: Decline to state & 478 & 0.05 & 0.06\\
\hline
Self-reported race: Hispanic (any other) & 984 & 0.11 & 0.11\\
\hline
Self-reported race: Other/multi-racial & 633 & 0.07 & 0.07\\
\hline
Self-reported race: White and no other & 6081 & 0.67 & 0.66\\
\hline
Self-reported party ID: Democrat & 4603 & 0.51 & 0.47\\
\hline
Self-reported party ID: Independent & 2118 & 0.23 & 0.24\\
\hline
Self-reported party ID: Other/NA & 423 & 0.05 & 0.05\\
\hline
Self-reported party ID: Republican & 1964 & 0.22 & 0.24\\
\hline
\end{tabular}
\label{tbl:descriptives_w3}
\end{table*}

\begin{table*}

\caption{Reported helpfulness of Birdwatch notes, Wave 3 (N=3,016, weighted by age and user state)}
\centering
\begin{tabular}[t]{l|r|r|r}
\hline
Response & \% agree & 5\% & 95\%\\
\hline
Extremely helpful & 32.71 & 30.72 & 34.76\\
\hline
Somewhat helpful & 31.47 & 29.48 & 33.53\\
\hline
Neither helpful nor unhelpful & 15.76 & 14.25 & 17.40\\
\hline
Somewhat unhelpful & 6.59 & 5.53 & 7.82\\
\hline
Extremely unhelpful & 13.47 & 12.03 & 15.06\\
\hline
\end{tabular}
\label{tbl:helpfulness_w3}
\end{table*}

Table \ref{tbl:w2_helpfulness} displays Wave 2 estimates of helpfulness weighted by demographic factors. 

\begin{table*}

\caption{Reported helpfulness of Birdwatch notes, Wave 2 (N=4,656, weighted)}
\centering
\footnotesize
\begin{tabular}[t]{l|rrr|rrr|rrr|l|rrr|rrr|rrr|l|rrr|rrr|rrr|l|rrr|rrr|rrr|l|rrr|rrr|rrr|l|rrr|rrr|rrr|l|rrr|rrr|rrr|l|rrr|rrr|rrr|l|rrr|rrr|rrr|l|rrr|rrr|rrr}
\hline
Response & Pct. (Any) & 5\% (Any) & 95\% (Any) & Pct. (Supermajority) & 5\% (Supermajority) & 95\% (Supermajority) & Pct. (MF) & 5\% (MF) & 95\% (MF)\\
\hline
Extremely helpful & 29.32 & 26.77 & 32.01 & 28.58 & 25.91 & 31.40 & 37.24 & 31.92 & 42.89\\
\hline
Somewhat helpful & 26.62 & 24.10 & 29.29 & 25.97 & 23.34 & 28.80 & 23.89 & 19.37 & 29.09\\
\hline
Neither helpful nor unhelpful & 17.06 & 14.92 & 19.44 & 17.50 & 15.20 & 20.06 & 17.28 & 13.37 & 22.03\\
\hline
Somewhat unhelpful & 6.21 & 4.94 & 7.77 & 6.26 & 4.93 & 7.92 & 5.06 & 3.02 & 8.37\\
\hline
Extremely unhelpful & 20.80 & 18.52 & 23.27 & 21.70 & 19.24 & 24.38 & 16.53 & 12.79 & 21.09\\
\hline
\end{tabular}
\label{tbl:w2_helpfulness}
\end{table*}

\begin{table*}

\caption{Pct. reporting Birdwatch notes ``at least somewhat helpful,'' by PID, Wave 2 (Weighted, N=4,656)}
\centering
\begin{tabular}[t]{c||r|r||r|r||r|r}
\hline
Party ID & Pct. (Any) & +/- (Any) & Pct. (Original) & +/- (Original) & Pcr. (Prototype) & +/- (Prototype)\\
\hline
Democrat & 72.05 & 3.24 & 71.27 & 3.47 & 74.96 & 5.66\\
\hline
Independent & 42.59 & 4.54 & 41.37 & 4.77 & 51.45 & 9.07\\
\hline
Other/NA & 44.32 & 8.92 & 39.29 & 9.30 & 48.73 & 17.04\\
\hline
Republican & 37.86 & 6.13 & 35.42 & 6.41 & 47.03 & 12.44\\
\hline
\end{tabular}
\label{tbl:helpful_bypid_w2}
\end{table*}

\begin{table*}

\caption{Pct. who say Birdwatch notes at least ``somewhat helpful,'' by party ID, Wave 3 (weighted by age and user state, N=3,016)}
\centering
\begin{tabular}[t]{l|r|r|r}
\hline
Party affiliation & \% & 5\% & 95\%\\
\hline
Democrat & 78.30 & 75.79 & 80.81\\
\hline
Independent & 52.82 & 48.47 & 57.18\\
\hline
Other/NA & 63.84 & 52.99 & 74.70\\
\hline
Republican & 48.22 & 43.55 & 52.90\\
\hline
\end{tabular}
\label{tbl:helpful_bypid_w3}
\end{table*}

\begin{table*}[!t] \centering 
  \caption{Models of the probability of agreeing with the substance of a potentially misleading Tweet in a survey experiment (Wave 2)} 
  \label{tbl:w2_tw_logit_models} 
\tiny 
\begin{tabular}{@{\extracolsep{0pt}}lD{.}{.}{-2} D{.}{.}{-2} D{.}{.}{-2} D{.}{.}{-2} D{.}{.}{-2} D{.}{.}{-2} } 
\\[-1.8ex]\hline 
\hline \\[-1.8ex] 
 & \multicolumn{6}{c}{\textit{Dependent variable:}} \\ 
\cline{2-7} 
\\[-1.8ex] & \multicolumn{6}{c}{Agree with substance of Tweet} \\ 
 & \multicolumn{2}{c}{Any} & \multicolumn{2}{c}{Supermajority Vote} & \multicolumn{2}{c}{Matrix Factorization} \\ 
\\[-1.8ex] & \multicolumn{1}{c}{(1)} & \multicolumn{1}{c}{(2)} & \multicolumn{1}{c}{(3)} & \multicolumn{1}{c}{(4)} & \multicolumn{1}{c}{(5)} & \multicolumn{1}{c}{(6)}\\ 
\hline \\[-1.8ex] 
 Treatment & -0.19^{**}$ $(0.07) & -0.10$ $(0.14) & -0.17^{*}$ $(0.08) & -0.20$ $(0.15) & -0.58^{***}$ $(0.14) & -0.38$ $(0.29) \\ 
  PID: Democrat &  & -0.45^{***}$ $(0.12) &  & -0.65^{***}$ $(0.13) &  & 0.03$ $(0.23) \\ 
  PID: Other/NA &  & 0.31$ $(0.20) &  & 0.29$ $(0.21) &  & 0.19$ $(0.40) \\ 
  PID: Republican &  & 0.61^{***}$ $(0.14) &  & 0.60^{***}$ $(0.15) &  & 0.17$ $(0.29) \\ 
  Age: 25-34 &  & -0.19$ $(0.11) &  & -0.20$ $(0.12) &  & -0.32$ $(0.21) \\ 
  Age: 35-44 &  & -0.20$ $(0.12) &  & -0.17$ $(0.13) &  & -0.38$ $(0.23) \\ 
  Age: 45-54 &  & -0.10$ $(0.12) &  & -0.10$ $(0.13) &  & -0.27$ $(0.23) \\ 
  Age: 55-64 &  & -0.10$ $(0.13) &  & -0.02$ $(0.14) &  & -0.63^{*}$ $(0.27) \\ 
  Age: 65+ &  & -0.13$ $(0.16) &  & -0.10$ $(0.18) &  & -0.37$ $(0.31) \\ 
  Gender: Non-binary &  & -0.11$ $(0.18) &  & -0.18$ $(0.19) &  & -0.26$ $(0.38) \\ 
  Gender: Other &  & 0.09$ $(0.27) &  & -0.02$ $(0.30) &  & 0.33$ $(0.48) \\ 
  Gender: Woman &  & -0.14$ $(0.08) &  & -0.13$ $(0.09) &  & -0.23$ $(0.16) \\ 
  Race: Asian (no other) &  & 0.17$ $(0.18) &  & 0.12$ $(0.20) &  & 0.36$ $(0.32) \\ 
  Race: Black (no other) &  & 0.42^{**}$ $(0.14) &  & 0.34^{*}$ $(0.15) &  & 0.92^{***}$ $(0.26) \\ 
  Race: Decline to state &  & 0.32^{*}$ $(0.15) &  & 0.32^{*}$ $(0.16) &  & -0.28$ $(0.34) \\ 
  Race: Hispanic (any other) &  & -0.09$ $(0.12) &  & 0.01$ $(0.13) &  & -0.45$ $(0.25) \\ 
  Race: Other/multi-racial &  & 0.13$ $(0.14) &  & 0.11$ $(0.15) &  & 0.11$ $(0.26) \\ 
  Treatment x PID: Democrat &  & -0.10$ $(0.18) &  & 0.07$ $(0.19) &  & -0.40$ $(0.36) \\ 
  Treatment x PID: Other/NA &  & -0.63^{*}$ $(0.32) &  & -0.47$ $(0.33) &  & -0.24$ $(0.59) \\ 
  Treatment x PID: Republican &  & 0.04$ $(0.21) &  & 0.24$ $(0.22) &  & -0.27$ $(0.44) \\ 
  Intercept & -1.18^{***}$ $(0.05) & -1.02^{***}$ $(0.13) & -1.23^{***}$ $(0.05) & -1.00^{***}$ $(0.14) & -0.77^{***}$ $(0.09) & -0.48$ $(0.25) \\ 
 \hline \\[-1.8ex] 
Observations & \multicolumn{1}{c}{4,512} & \multicolumn{1}{c}{4,512} & \multicolumn{1}{c}{4,027} & \multicolumn{1}{c}{4,027} & \multicolumn{1}{c}{1,081} & \multicolumn{1}{c}{1,081} \\ 
Log Likelihood & \multicolumn{1}{c}{-2,366.70} & \multicolumn{1}{c}{-2,280.89} & \multicolumn{1}{c}{-2,074.19} & \multicolumn{1}{c}{-1,975.78} & \multicolumn{1}{c}{-615.02} & \multicolumn{1}{c}{-597.32} \\ 
Akaike Inf. Crit. & \multicolumn{1}{c}{4,737.41} & \multicolumn{1}{c}{4,603.77} & \multicolumn{1}{c}{4,152.38} & \multicolumn{1}{c}{3,993.56} & \multicolumn{1}{c}{1,234.05} & \multicolumn{1}{c}{1,236.64} \\ 
\hline 
\hline \\[-1.8ex] 
\textit{Note:}  & \multicolumn{6}{r}{$^{*}$p$<$0.05; $^{**}$p$<$0.01; $^{***}$p$<$0.001} \\ 
 & \multicolumn{6}{r}{All models are logit, unweighted, estimated using maximum likelihood.} \\ 
 & \multicolumn{6}{r}{Reference category is politically independent, white, male, aged 18-24.} \\ 
 & \multicolumn{6}{r}{Coefficients are the natural logarithms of odds ratios.} \\ 
\end{tabular} 
\end{table*}

\begin{table*}[!t] \centering 
  \caption{Models of the probability of agreeing with the substance of a note on a potentially misleading Tweet in a survey experiment (Wave 2)} 
  \label{tbl:w2_nt_logit_models} 
\tiny 
\begin{tabular}{@{\extracolsep{0pt}}lD{.}{.}{-2} D{.}{.}{-2} D{.}{.}{-2} } 
\\[-1.8ex]\hline 
\hline \\[-1.8ex] 
 & \multicolumn{3}{c}{\textit{Dependent variable:}} \\ 
\cline{2-4} 
\\[-1.8ex] & \multicolumn{3}{c}{Agree with substance of note} \\ 
 & \multicolumn{1}{c}{Any} & \multicolumn{1}{c}{Supermajority Vote} & \multicolumn{1}{c}{Matrix Factorization} \\ 
\\[-1.8ex] & \multicolumn{1}{c}{(1)} & \multicolumn{1}{c}{(2)} & \multicolumn{1}{c}{(3)}\\ 
\hline \\[-1.8ex] 
 Treatment & 0.55^{***}$ $(0.12) & 0.56^{***}$ $(0.13) & 0.96^{***}$ $(0.26) \\ 
  PID: Democrat & 0.44^{***}$ $(0.10) & 0.52^{***}$ $(0.11) & 0.32$ $(0.22) \\ 
  PID: Other/NA & -0.46^{*}$ $(0.20) & -0.36$ $(0.21) & -0.26$ $(0.41) \\ 
  PID: Republican & 0.10$ $(0.13) & 0.09$ $(0.14) & 0.41$ $(0.28) \\ 
  Age: 25-34 & 0.20^{*}$ $(0.10) & 0.20$ $(0.10) & 0.10$ $(0.19) \\ 
  Age: 35-44 & 0.06$ $(0.10) & 0.02$ $(0.11) & -0.06$ $(0.21) \\ 
  Age: 45-54 & 0.05$ $(0.10) & 0.04$ $(0.11) & 0.19$ $(0.21) \\ 
  Age: 55-64 & -0.21$ $(0.11) & -0.22$ $(0.12) & -0.06$ $(0.23) \\ 
  Age: 65+ & -0.15$ $(0.14) & -0.15$ $(0.15) & -0.27$ $(0.28) \\ 
  Gender: Non-binary & 0.44^{**}$ $(0.15) & 0.51^{**}$ $(0.16) & 0.39$ $(0.34) \\ 
  Gender: Other & -0.58^{*}$ $(0.25) & -0.66^{*}$ $(0.26) & -1.19^{*}$ $(0.54) \\ 
  Gender: Woman & 0.002$ $(0.07) & -0.001$ $(0.07) & -0.10$ $(0.14) \\ 
  Race: Asian (no other) & -0.53^{***}$ $(0.16) & -0.61^{***}$ $(0.17) & -0.42$ $(0.30) \\ 
  Race: Black (no other) & -0.67^{***}$ $(0.12) & -0.63^{***}$ $(0.13) & -0.40$ $(0.26) \\ 
  Race: Decline to state & -0.37^{**}$ $(0.14) & -0.35^{*}$ $(0.14) & 0.33$ $(0.29) \\ 
  Race: Hispanic (any other) & -0.35^{***}$ $(0.10) & -0.38^{***}$ $(0.10) & -0.05$ $(0.20) \\ 
  Race: Other/multi-racial & -0.30^{*}$ $(0.12) & -0.28^{*}$ $(0.13) & -0.27$ $(0.24) \\ 
  Treatment x PID: Democrat & 0.30^{*}$ $(0.15) & 0.30$ $(0.16) & 0.06$ $(0.31) \\ 
  Treatment x PID: Other/NA & -0.05$ $(0.28) & -0.20$ $(0.30) & -0.02$ $(0.55) \\ 
  Treatment x PID: Republican & -0.55^{**}$ $(0.19) & -0.70^{***}$ $(0.21) & -0.48$ $(0.39) \\ 
  Intercept & -0.20$ $(0.11) & -0.23$ $(0.12) & -0.57^{*}$ $(0.24) \\ 
 \hline \\[-1.8ex] 
Observations & \multicolumn{1}{c}{4,512} & \multicolumn{1}{c}{4,027} & \multicolumn{1}{c}{1,081} \\ 
Log Likelihood & \multicolumn{1}{c}{-2,939.97} & \multicolumn{1}{c}{-2,602.64} & \multicolumn{1}{c}{-711.02} \\ 
Akaike Inf. Crit. & \multicolumn{1}{c}{5,921.94} & \multicolumn{1}{c}{5,247.29} & \multicolumn{1}{c}{1,464.03} \\ 
\hline 
\hline \\[-1.8ex] 
\textit{Note:}  & \multicolumn{3}{r}{$^{*}$p$<$0.05; $^{**}$p$<$0.01; $^{***}$p$<$0.001} \\ 
 & \multicolumn{3}{r}{All models are logit, unweighted, estimated using maximum likelihood.} \\ 
 & \multicolumn{3}{r}{Reference category is politically independent, white, male, aged 18-24.} \\ 
 & \multicolumn{3}{r}{Coefficients are the natural logarithms of odds ratios.} \\ 
\end{tabular} 
\end{table*}

\begin{table*}[!t] \centering 
  \caption{Models of the probability of agreeing with the substance of a potentially misleading Tweet and associated note in a survey experiment (Wave 3)} 
  \label{tbl:w3_logit_models} 
\small 
\begin{tabular}{@{\extracolsep{2pt}}lD{.}{.}{-2} D{.}{.}{-2} D{.}{.}{-2} D{.}{.}{-2} } 
\\[-1.8ex]\hline 
\hline \\[-1.8ex] 
 & \multicolumn{4}{c}{\textit{Dependent variable:}} \\ 
\cline{2-5} 
\\[-1.8ex] & \multicolumn{2}{c}{Agree with substance of Tweet} & \multicolumn{2}{c}{Agree with substance of note} \\ 
\\[-1.8ex] & \multicolumn{1}{c}{(1)} & \multicolumn{1}{c}{(2)} & \multicolumn{1}{c}{(3)} & \multicolumn{1}{c}{(4)}\\ 
\hline \\[-1.8ex] 
 Treatment & -0.46^{***}$ $(0.13) & -0.46^{***}$ $(0.13) & 0.90^{***}$ $(0.11) & 0.91^{***}$ $(0.11) \\ 
  PID: Democrat & 0.29^{**}$ $(0.10) & 0.31^{**}$ $(0.10) & 0.12$ $(0.10) & 0.14$ $(0.10) \\ 
  PID: Other/NA & -0.20$ $(0.21) & -0.25$ $(0.21) & -0.38$ $(0.20) & -0.35$ $(0.20) \\ 
  PID: Republican & 0.01$ $(0.12) & 0.05$ $(0.12) & 0.04$ $(0.11) & 0.04$ $(0.11) \\ 
  Age: 18-29 &  & 0.20^{*}$ $(0.09) &  & -0.004$ $(0.08) \\ 
  Age: 40-49 &  & -0.03$ $(0.09) &  & -0.08$ $(0.08) \\ 
  Age: 50-59 &  & -0.11$ $(0.10) &  & -0.15$ $(0.09) \\ 
  Age: 60+ &  & 0.03$ $(0.11) &  & -0.23^{*}$ $(0.10) \\ 
  Gender: Non-binary &  & 0.02$ $(0.16) &  & 0.22$ $(0.15) \\ 
  Gender: Other &  & 0.10$ $(0.24) &  & 0.14$ $(0.22) \\ 
  Gender: Woman &  & -0.09$ $(0.07) &  & -0.09$ $(0.06) \\ 
  Race: Asian (no other) &  & 0.01$ $(0.17) &  & -0.25$ $(0.15) \\ 
  Race: Black (no other) &  & 0.22$ $(0.12) &  & -0.35^{**}$ $(0.11) \\ 
  Race: Decline to state &  & -0.14$ $(0.15) &  & -0.31^{*}$ $(0.13) \\ 
  Race: Hispanic (any other) &  & 0.09$ $(0.10) &  & -0.05$ $(0.09) \\ 
  Race: Other/multi-racial &  & 0.06$ $(0.12) &  & -0.09$ $(0.11) \\ 
  Treatment x PID: Democrat & -0.10$ $(0.16) & -0.11$ $(0.16) & 0.54^{***}$ $(0.14) & 0.54^{***}$ $(0.14) \\ 
  Treatment x PID: Other/NA & 0.18$ $(0.32) & 0.23$ $(0.32) & 0.34$ $(0.27) & 0.32$ $(0.27) \\ 
  Treatment x PID: Republican & 0.04$ $(0.19) & 0.05$ $(0.19) & -0.16$ $(0.16) & -0.15$ $(0.16) \\ 
  Intercept & -1.01^{***}$ $(0.09) & -1.06^{***}$ $(0.11) & -0.62^{***}$ $(0.08) & -0.48^{***}$ $(0.10) \\ 
 \hline \\[-1.8ex] 
Observations & \multicolumn{1}{c}{6,017} & \multicolumn{1}{c}{6,017} & \multicolumn{1}{c}{6,017} & \multicolumn{1}{c}{6,017} \\ 
Log Likelihood & \multicolumn{1}{c}{-3,348.96} & \multicolumn{1}{c}{-3,337.30} & \multicolumn{1}{c}{-3,886.17} & \multicolumn{1}{c}{-3,870.45} \\ 
Akaike Inf. Crit. & \multicolumn{1}{c}{6,713.91} & \multicolumn{1}{c}{6,714.60} & \multicolumn{1}{c}{7,788.34} & \multicolumn{1}{c}{7,780.90} \\ 
\hline 
\hline \\[-1.8ex] 
\textit{Note:}  & \multicolumn{4}{r}{$^{*}$p$<$0.05; $^{**}$p$<$0.01; $^{***}$p$<$0.001} \\ 
 & \multicolumn{4}{r}{All models are logit, unweighted, estimated using maximum likelihood.} \\ 
 & \multicolumn{4}{r}{Reference category is politically independent, white, male, aged 30-39.} \\ 
 & \multicolumn{4}{r}{Coefficients are the natural logarithms of odds ratios.} \\ 
\end{tabular} 
\end{table*} 

\begin{table*}[!t] \centering 
  \caption{Model of Tweet agreement on five-point scale, Wave 3 (for comparison to fact-checks literature).} 
  \label{tbl:w3_tw_5pt_model} 
\small 
\begin{tabular}{@{\extracolsep{5pt}}lD{.}{.}{-2} } 
\\[-1.8ex]\hline 
\hline \\[-1.8ex] 
 & \multicolumn{1}{c}{\textit{Dependent variable:}} \\ 
\cline{2-2} 
\\[-1.8ex] & \multicolumn{1}{c}{Tweet agreement (5-point scale)} \\ 
\hline \\[-1.8ex] 
 Treatment & -0.46^{***}$ $(0.09) \\ 
  PID: Democrat & 0.13$ $(0.08) \\ 
  PID: Other/NA & -0.03$ $(0.17) \\ 
  PID: Republican & -0.08$ $(0.09) \\ 
  Age: 18-29 & 0.18^{**}$ $(0.06) \\ 
  Age: 40-49 & 0.01$ $(0.07) \\ 
  Age: 50-59 & -0.07$ $(0.07) \\ 
  Age: 60+ & 0.02$ $(0.08) \\ 
  Gender: Non-binary & 0.09$ $(0.12) \\ 
  Gender: Other & 0.03$ $(0.18) \\ 
  Gender: Woman & 0.04$ $(0.05) \\ 
  Race: Asian (no other) & -0.02$ $(0.12) \\ 
  Race: Black (no other) & 0.18^{*}$ $(0.09) \\ 
  Race: Decline to state & 0.02$ $(0.11) \\ 
  Race: Hispanic (any other) & 0.09$ $(0.07) \\ 
  Race: Other/multi-racial & 0.10$ $(0.09) \\ 
  Treatment x PID: Democrat & -0.19$ $(0.11) \\ 
  Treatment x PID: Other/NA & 0.14$ $(0.23) \\ 
  Treatment x PID: Republican & 0.07$ $(0.13) \\ 
  Intercept & -0.38^{***}$ $(0.08) \\ 
 \hline \\[-1.8ex] 
Observations & \multicolumn{1}{c}{4,948} \\ 
R$^{2}$ & \multicolumn{1}{c}{0.04} \\ 
Adjusted R$^{2}$ & \multicolumn{1}{c}{0.03} \\ 
Residual Std. Error & \multicolumn{1}{c}{1.54 (df = 4928)} \\ 
F Statistic & \multicolumn{1}{c}{9.97$^{***}$ (df = 19; 4928)} \\ 
\hline 
\hline \\[-1.8ex] 
\textit{Note:}  & \multicolumn{1}{r}{$^{*}$p$<$0.05; $^{**}$p$<$0.01; $^{***}$p$<$0.001} \\ 
 & \multicolumn{1}{r}{Unweighted OLS. Scale on interval [-2,2], where lower values indicate less agreement.} \\ 
 & \multicolumn{1}{r}{Reference category is politically independent, white, male, aged 30-39.} \\ 
\end{tabular} 
\end{table*}

\begin{table*}[!t] \centering 
  \caption{Models of the probability of appraising a Tweet as ``at least somewhat accurate'' in a survey experiment (Wave 3)} 
  \label{tbl:w3_logit_acc} 
\small 
\begin{tabular}{@{\extracolsep{2pt}}lD{.}{.}{-2} D{.}{.}{-2} } 
\\[-1.8ex]\hline 
\hline \\[-1.8ex] 
 & \multicolumn{2}{c}{\textit{Dependent variable:}} \\ 
\cline{2-3} 
\\[-1.8ex] & \multicolumn{2}{c}{Appraise potentially misleading Tweet as ``at least somewhat accurate''} \\ 
\\[-1.8ex] & \multicolumn{1}{c}{(1)} & \multicolumn{1}{c}{(2)}\\ 
\hline \\[-1.8ex] 
 Treatment & 0.06$ $(0.14) & 0.07$ $(0.14) \\ 
  PID: Democrat & 0.66^{***}$ $(0.12) & 0.63^{***}$ $(0.12) \\ 
  PID: Other/NA & -0.28$ $(0.26) & -0.30$ $(0.26) \\ 
  PID: Republican & 0.26$ $(0.14) & 0.23$ $(0.14) \\ 
  Age: 18-29 &  & 0.08$ $(0.09) \\ 
  Age: 40-49 &  & 0.04$ $(0.10) \\ 
  Age: 50-59 &  & 0.01$ $(0.10) \\ 
  Age: 60+ &  & 0.45^{***}$ $(0.11) \\ 
  Gender: Non-binary &  & -0.21$ $(0.18) \\ 
  Gender: Other &  & -0.43$ $(0.30) \\ 
  Gender: Woman &  & -0.07$ $(0.07) \\ 
  Race: Asian (no other) &  & -0.04$ $(0.18) \\ 
  Race: Black (no other) &  & 0.26^{*}$ $(0.12) \\ 
  Race: Decline to state &  & -0.02$ $(0.16) \\ 
  Race: Hispanic (any other) &  & 0.13$ $(0.10) \\ 
  Race: Other/multi-racial &  & 0.08$ $(0.13) \\ 
  Treatment x PID: Democrat & -0.34^{*}$ $(0.17) & -0.34^{*}$ $(0.17) \\ 
  Treatment x PID: Other/NA & 0.06$ $(0.36) & 0.08$ $(0.36) \\ 
  Treatment x PID: Republican & -0.13$ $(0.20) & -0.13$ $(0.20) \\ 
  Intercept & -1.58^{***}$ $(0.10) & -1.66^{***}$ $(0.12) \\ 
 \hline \\[-1.8ex] 
Observations & \multicolumn{1}{c}{6,017} & \multicolumn{1}{c}{6,017} \\ 
Log Likelihood & \multicolumn{1}{c}{-3,152.04} & \multicolumn{1}{c}{-3,136.49} \\ 
Akaike Inf. Crit. & \multicolumn{1}{c}{6,320.08} & \multicolumn{1}{c}{6,312.97} \\ 
\hline 
\hline \\[-1.8ex] 
\textit{Note:}  & \multicolumn{2}{r}{$^{*}$p$<$0.05; $^{**}$p$<$0.01; $^{***}$p$<$0.001} \\ 
 & \multicolumn{2}{r}{All models are logit, unweighted, estimated using maximum likelihood.} \\ 
 & \multicolumn{2}{r}{Reference category is politically independent, white, male, aged 30-39.} \\ 
 & \multicolumn{2}{r}{Coefficients are the natural logarithms of odds ratios.} \\ 
\end{tabular} 
\end{table*} 

\clearpage
\newpage

\section{Appendix: Note Selection} \label{note_selection}

For Wave 2, we began note selection with 202 notes from the preceding two-week span (Nov. 1--Nov.13, 2021).

Each note was scored by six algorithms, and any note achieving a high enough score on any of the algorithms was eligible to be in the sample. The algorithms included a weighted average helpfulness score, an author-rater reputation-based score, a cross-partisan (crossover score), a score based on contributor-generated note tags, and the remaining were matrix factorization-based algorithms. The top 20\% of notes generated by MF algorithms and the tag-based algorithm were eligible for inclusion. The remaining algorithms were binary, and any note receiving a nonzero score was eligible to be included. This procedure reduced the set from 202 to 89 eligible notes (44\%).  

Next, fourteen notes were removed because they contained a ``not misleading'' tag, as the core focus of the study were notes that were potentially misleading. Fourteen notes excluded from analysis because notes on “not misleading” Tweets are not a current focus of study.

To exclude the possibility of including any harmful misinformation in the study, five notes were excluded because they applied to Tweets that appeared to be harmful misinformation or to Tweets that we suspected may have violated Twitter policy.

We also excluded notes on content that does not make sense out of context. Twenty-four additional notes were excluded because they do not make sense out of the context of Twitter at a specific time. This included notes on Tweets that have been deleted; notes on Tweets that only make sense in the context of a larger Twitter thread or of a specific and transitory topic of discussion on Twitter; notes that would require the reader to reference an embedded video or follow a link in order to really evaluate the note; notes on Tweets that don’t make sense unless the reader can inspect embedded media that would not be legible in the survey context, including Tweets with multiple embedded photos (in this context the photos may render too small to be visible)

Finally, we removed three potentially misleading notes. These notes were removed because they contain errors of fact regarding matters of broad public interest, and for ethical reasons we had a strict policy against misinforming survey participants.

A similar approach was used to select notes for Wave 3, except a single algorithm was used to determine initial eligibility for inclusion. We began note selection with 103 notes from the preceding four-week span (March 21, 2022). Each note was scored by the MF algorithm, and any note achieving a high enough score on the algorithm (.3 or higher) was eligible to be in the sample. This represented the top 6\% of notes at the time, but it also included a number of notes below the threshold above which notes were eligible to receive ``currently rated helpful'' status (0.4). Twenty-six notes were rendered ineligible because they were on deleted Tweets. We also excluded 17 notes on content that would not make sense out of context; 11 notes on Tweets that we suspected included harmful misinformation or that may violate Twitter policy; and three notes that contained errors of fact on matters of broad public interest.

\clearpage
\newpage

\section{Appendix: Experiment Details} \label{experiment_details}

Prior to expanding viewership to all of US (early October), we had been running A/B tests in order to understand the impact that Birdwatch notes have on engagement and sharing behavior on regular users. Between July 27th and September 27th, 2022, 20\% of US users were eligible to see currently rated helpful (CRH) notes placed below the tweet. 
We observe statistically significant changes (with 95\% CI) on engagement rates across several forms of interaction. Some of the most consequential, also outlined in Tables \ref{tab:exp_deltas} and \ref{tab:exp_raw}, include favorite, retweet, and quote actions. These results corroborate both our original hypothesis as well as survey outcomes outlined above. 

Over the course of the experiment, approx 2.4M users in the test group had at least one linger impression on a tweet with a Birdwatch note. Measuring behavioral changes across multiple 14 day observation windows, the outcomes are consistent over the duration of the experiment (see Fig. \ref{fig:exp_results}). 
Each 14 day window covers between 2M and 6.8M tweet impressions. Note that this variance is due to exogenous factors (ongoing events, number of visible notes,  tweet distribution etc), as experiment conditions remain fixed.
 \\
We use the following steps to estimate the difference in engagement rates between test and control for each interaction. Computing the change in retweet rate, for example, we define the following quantities:   

\noindent
$ p_{retweet} = \frac{retweets}{impressions} $ \\

\noindent
$ \text {change in retweet rate} = \frac{p_{retweet_{test}}}{p_{retweet_{control}}}-1$ \\

\noindent
Assuming a binomial distribution, the standard error (SE), in absolute terms, follows:\\ 
\noindent
$ SE_{retweet} = \sqrt(\frac{1}{impressions_{test}} + \frac{1}{impressions_{control}}) * pooled p_{retweet} * (1 - pooled p_{retweet}) \text{, where}$ \\

$ pooled p_{retweet} = \frac{(retweets_{test} + retweets_{control})}{(impressions_{test} + impressions_{control})} $ \\

\begin{table}[ht]
\caption{A/B test relative \% difference in engagement rates}
\centering
\begin{tabular}{rlrrr}
  \hline
 & Time and Interaction Type & \% Difference & 95\% CI Lower & 95\% CI Upper \\ 
  \hline
1 & T1 favorite & -34.06 & -36.54 & -31.58 \\ 
  2 & T1 retweet & -29.78 & -34.97 & -24.59 \\ 
  3 & T1 quote & -14.15 & -24.95 & -3.34 \\ 
  4 & T1 follow & -11.37 & -72.60 & 49.85 \\ 
  5 & T2 favorite & -32.75 & -33.98 & -31.53 \\ 
  6 & T2 retweet & -28.78 & -31.28 & -26.28 \\ 
  7 & T2 quote & -26.51 & -33.65 & -19.36 \\ 
  8 & T2 follow & -6.29 & -36.81 & 24.23 \\ 
  9 & T3 favorite & -28.36 & -29.75 & -26.97 \\ 
  10 & T3 retweet & -28.10 & -31.25 & -24.95 \\ 
  11 & T3 quote & -21.43 & -29.64 & -13.23 \\ 
  12 & T3 follow & -22.39 & -54.83 & 10.05 \\ 
   \hline
\end{tabular}
{\caption*{Relative difference between test and control as measured across three different time-windows; T1: Aug 7th - Aug 20th, T2: Aug 21st - Sept 3rd, T3: Sept 4th - Sept 17th} } 
\label{tab:exp_deltas}
\end{table}

\clearpage

\begin{table}[ht]
\caption{A/B test underlying data}
\centering
\begin{tabular}{rlrrrrr}
  \hline
 & Time Window and Interaction & Count (Test) & Count (Control) & Rate (Test) & Rate (Control) & Standard Error \\ 
  \hline
1 & T1 favorite & 9644 & 15009 & 0.0096 & 0.0145 & 0.000153 \\ 
  2 & T1 retweet & 2310 & 3376 & 0.0023 & 0.0033 & 0.000074 \\ 
  3 & T1 quote & 599 & 716 & 0.0006 & 0.0007 & 0.000036 \\ 
  4 & T1 follow & 19 & 22 & 0.0000 & 0.0000 & 0.000006 \\ 
  5 & T2 favorite & 40154 & 60914 & 0.0119 & 0.0176 & 0.000092 \\ 
  6 & T2 retweet & 10081 & 14440 & 0.0030 & 0.0042 & 0.000046 \\ 
  7 & T2 quote & 1260 & 1749 & 0.0004 & 0.0005 & 0.000016 \\ 
  8 & T2 follow & 79 & 86 & 0.0000 & 0.0000 & 0.000004 \\ 
  9 & T3 favorite & 32022 & 45703 & 0.0156 & 0.0218 & 0.000133 \\ 
  10 & T3 retweet & 6376 & 9067 & 0.0031 & 0.0043 & 0.000060 \\ 
  11 & T3 quote & 992 & 1291 & 0.0005 & 0.0006 & 0.000023 \\ 
  12 & T3 follow & 63 & 83 & 0.0000 & 0.0000 & 0.000006 \\ 
   \hline
\end{tabular}
 {\caption*{Impressions T1: 1,006,245(test), 1,032,634(control); T2: 3,386,874(test), 3,455,092(control); T3: 2,052,516(test), 2,098,654(control)}}
\label{tab:exp_raw}
\end{table}

\clearpage
\newpage

\clearpage
\newpage

\section{Appendix: Product Screenshots} \label{product_screenshots}

\begin{figure}[h]
    \centering
    \includegraphics[width=0.5\columnwidth, trim={10px 0 0 0},clip]{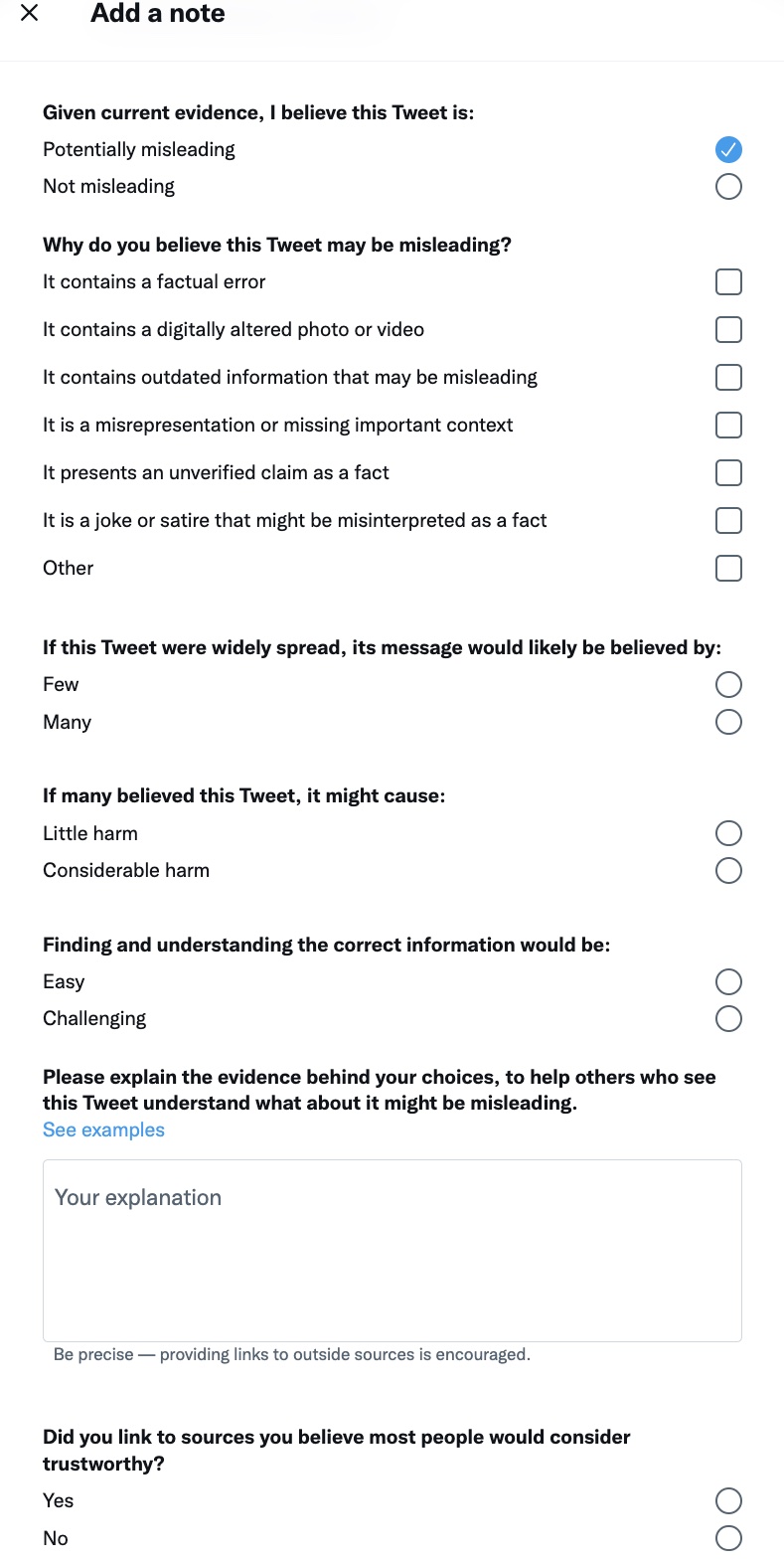}
    \caption{Note Creation Form.}
    \label{fig:note_form}
\end{figure}

\begin{figure}[h]
    \centering
    \includegraphics[width=0.5\columnwidth, trim={10px 0 0 0},clip]{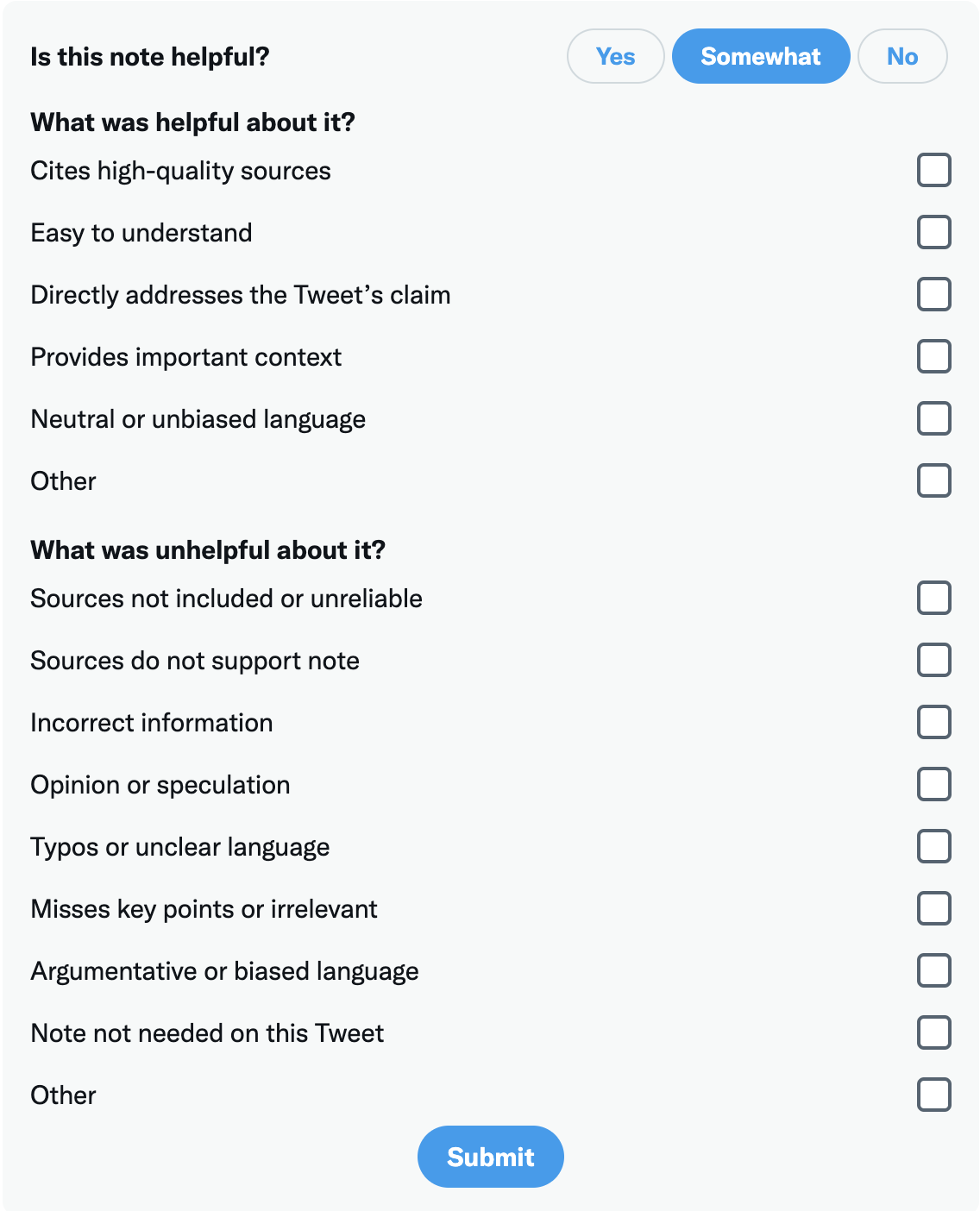}
    \caption{Rating Creation Form.}
    \label{fig:rating_form}
\end{figure}

\begin{figure}[h]
    \centering
    \includegraphics[width=0.5\columnwidth, trim={10px 0 0 0},clip]{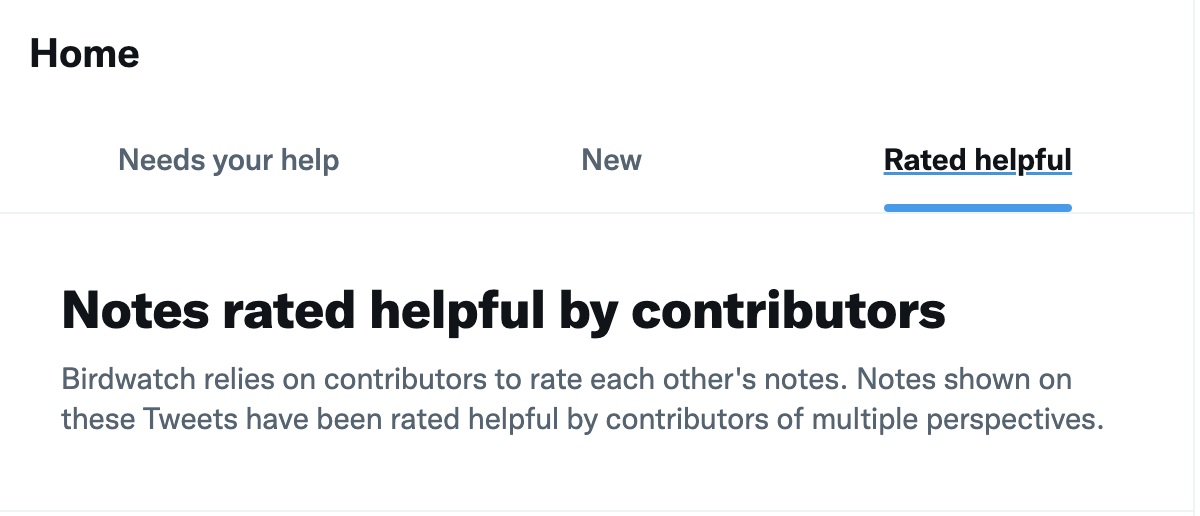}
    \caption{The tabs of the Birdwatch home page.}
    \label{fig:bw_home}
\end{figure}

\begin{figure}[h]
    \centering
    \includegraphics[width=0.5\columnwidth, trim={10px 0 0 0},clip]{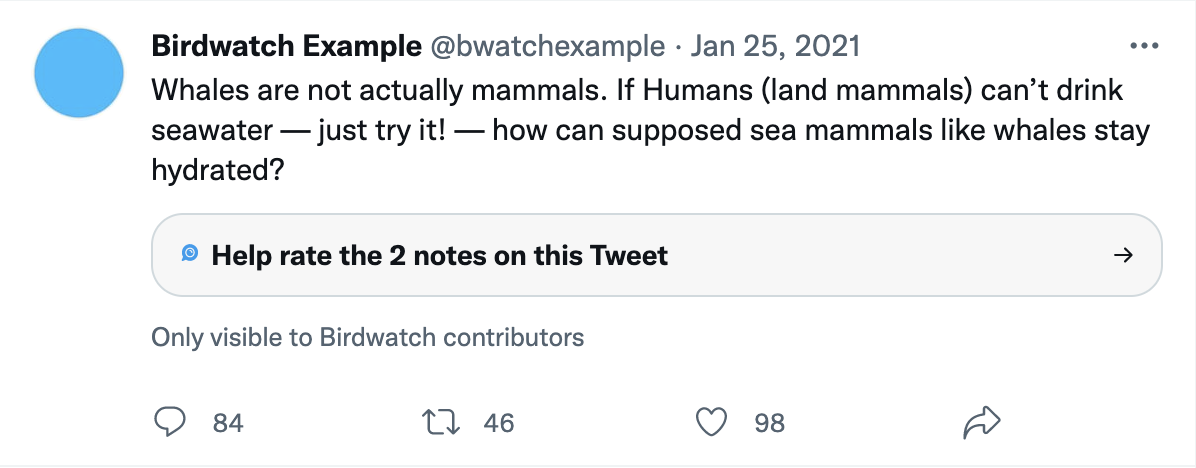}
    \caption{How a Tweet looks to a Birdwatch contributor when the Tweet has notes that need ratings, but has no currently rated helpful notes.}
    \label{fig:tweet_screenshot}
\end{figure}

Notes and Ratings are created by users who were using the Twitter Birdwatch product. Here, we show screenshots of the note and rating creation forms (Figure \ref{fig:note_form} and Figure \ref{fig:rating_form}), and two views of how contributors might find notes to rate (Figure \ref{fig:bw_home} and Figure \ref{fig:tweet_screenshot}).

\clearpage

\clearpage 

\end{document}